\pdfoutput=1
\documentclass[prb,reprint, epsfig,amssymb,bibliography,amsmath,footinbib,superscriptaddress]{revtex4-2}

\usepackage{blkarray}
\usepackage[dvipsnames]{xcolor}

\usepackage{amssymb,amsmath,bm,epsfig,graphicx,subfigure,hyperref}

\usepackage[utf8]{inputenc}
\usepackage[vietnamese, english]{babel}
\usepackage{chngcntr}
\usepackage{xcolor}
\usepackage{braket}
\usepackage{physics}
\usepackage{sidecap}
\usepackage{lipsum}
\usepackage{amsfonts}
\usepackage{bbold}
\usepackage{tensor}
\usepackage{makecell}
\usepackage{multirow}

\begin{document}

\title{\color{Blue}{Optical Absorption and Emission from Wannier-Stark Spectra of Moir\'{e} Superlattices}}
\author{\foreignlanguage{vietnamese}{Võ Tiến Phong}}
\email{vophong@magnet.fsu.edu}
\affiliation{Department of Physics, Florida State University, Tallahassee, Florida, 32306, USA}
\affiliation{National High Magnetic Field Laboratory, Tallahassee, Florida, 32310, USA}
\author{Francisco Guinea}
\email{paco.guinea@imdea.org}
\affiliation{IMDEA Nanoscience, Faraday 9, 28049 Madrid, Spain}
\affiliation{Donostia International Physics Center, Paseo Manuel de Lardiz\'{a}bal 4, 20018 San Sebasti\'{a}n, Spain}
\author{Cyprian Lewandowski}
\email{clewandowski@magnet.fsu.edu}
\affiliation{Department of Physics, Florida State University, Tallahassee, Florida, 32306, USA}
\affiliation{National High Magnetic Field Laboratory, Tallahassee, Florida, 32310, USA}

\date{May 20, 2025}

\begin{abstract} 
Using a formalism based on the non-Abelian Berry connection, we explore quantum geometric signatures of Wannier-Stark spectra in two-dimensional superlattices. The Stark energy can be written as \textit{intraband} Berry phases, while Zener tunneling is given by \textit{interband} Berry connections. We suggest that the gaps induced by interband hybridization can be probed by THz optical absorption and emission spectroscopy. This is especially relevant to modern moir\'{e} materials wherein mini-bands are often spectrally entangled, leading to strong interband hybridization in the Wannier-Stark regime.  Furthermore, owing to their large superlattice constants, both the low-field and high-field regimes can be accessed in these materials using presently available technology. Importantly, even at moderate electric fields, we find that stimulated emission can dominate absorption, raising the possibility of lasing at practically relevant parameter regimes.
\end{abstract}

\maketitle

\section{Introduction}

\begin{figure}[h]
    \centering
    \includegraphics[width=3.4in]{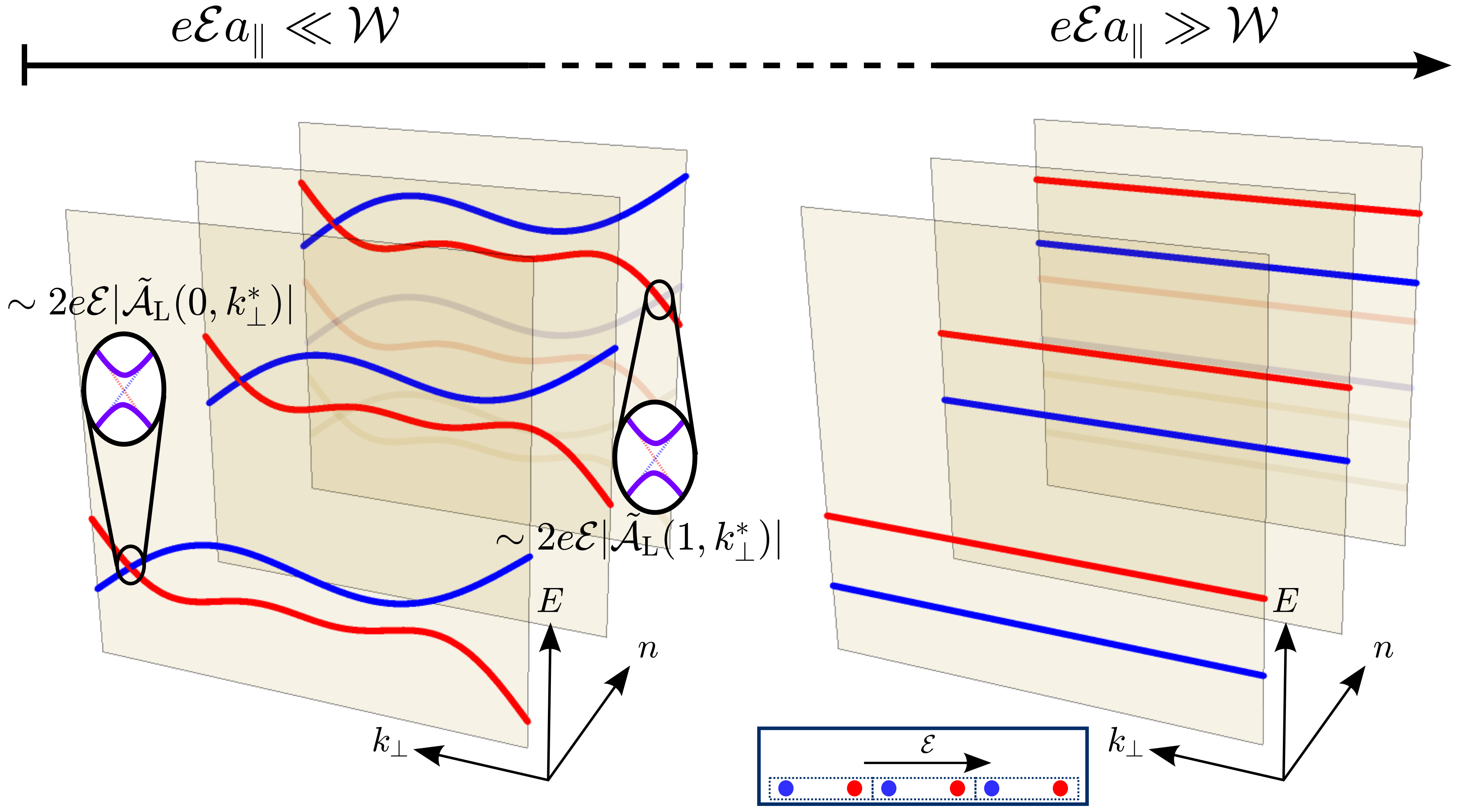}
    \caption{\textbf{Schematic representation of Wannier-Stark spectra of topological bands in two limiting cases of applied electric field.} The three-dimensional axes are energy $E$, crystal momentum in the direction perpendicular to the electric field $k_\perp,$ and Wannier-Stark ladder index along the direction parallel to the electric field $n.$ In regimes of small fields $e\mathcal{E}a_\parallel \ll \mathcal{W}$ (left panel), the decoupled Wannier-Stark bands capture the essence of the band structure. Here, the bands are topological because they wind around $k_\perp.$ Intersections must therefore occur, which are  gapped out by interband Zener tunneling. The magnitudes of these gaps are given approximately by the Fourier transform of the Berry connection in the ladder representation. In the $e\mathcal{E}a_\parallel \gg \mathcal{W}$ limit (right panel), the Stark energy on localized orbitals dominates the physics. The inter-orbital hoppings become insignificant, as shown in the inset. So the resulting Wannier-Stark bands are dispersionless.}
    \label{fig:schematic_figure}
\end{figure}

Quantum  electrons in a crystalline solid subjected to a   homogeneous, time-independent electric field display a surprisingly rich phenomenology of both fundamental interest and practical importance \cite{Wannier1962Dynamics, Nenciu1991Dynamics}. In the time domain, under ideal conditions, these electrons  execute Bloch oscillations with frequency $\omega_B = e\mathcal{E}a/\hbar,$ where $\mathcal{E}$ is the field magnitude and $a$ is the crystal period \cite{bloch1929quantenmechanik, zener1934theory}. In the frequency domain, the electric field  fractures   otherwise continuous energy bands indexed by crystal momentum into discrete Wannier-Stark ladders indexed by lattice positions with ladder rungs spectrally separated from each other by multiples of $\hbar \omega_B$ (a two-dimensional example is shown in Fig. \ref{fig:schematic_figure}) \cite{James1949Electronic,Katsura1950On,Feuer1952Electronic, Wannier1960Wave}.  Despite their early formulations, these concepts remained purely theoretical for many decades because their potential experimental realizations in natural crystals at detectable values of $\omega_B$ demanded, in part, insurmountably high electric fields. This stringent requirement was ingeniously circumvented by the introduction of engineered semiconductor quasi-one-dimensional superlattices wherein the lattice constant can be enlarged to considerably reduce the required field strengths \cite{mendez1993wannier, Rueda1995Wannier, Kurz1996Bloch, Nakayama1996Wannier, leo1998interband, Rossi1998Bloch, Sudzius2001Coherent, Meier2001Coherent}. This breakthrough led to numerous observations of signatures of Bloch oscillations and Wannier-Stark ladders using  complementary experimental techniques: photocurrent \cite{Mendez1988Stark, fujiwara1989room,Soucail1990Electron, kawashima1991room, gibb1993photocurrent},  emission spectroscopy \cite{Waschke1993Coherent, unuma2021room}, four-wave mixing \cite{Plessen1992Method, leo1992observation, Feldmann1992Optical, Leisching1994Bloch}, electro-optics \cite{Dekorsy1994Electro,Sekine2005Dispersive}, reflectance or transmission spectroscopy \cite{Voisin1988Observation,gibb1993photocurrent, Gibb1993Observation, Cho1996Bloch}, and  steady-state transport \cite{esaki1970superlattice,Sibille1990Observation,Beltram1990Scattering, Grahn1991Electrical,Tsu1991Stark}. Several of these experiments were even done at room temperature \cite{fujiwara1989room, mendez1990temperature, kawashima1991room, Dekorsy1995Bloch,dragoman2021bloch, unuma2021room}. Sustained interest in these coherent phenomena over many decades is rooted in their promise for potential future electronic devices such as tunable THz emitters \cite{Kurz1996Bloch,leo1998interband,karl2017basics}. While quasi-one-dimensional semiconductor superlattices provide proof-of-concept demonstrations of the fundamental idea, modern material platforms enable new physics stemming from macroscopic two-dimensional superlattice structures endowed with an unprecedented level of device tunability, motivating us to examine the Wannier-Stark regime in moir\'{e} materials.

In recent years, moir\'{e} superlattices have been constructed via interlayer interference controlled by twist angle, lattice constant mismatch, and/or strain \cite{geim2013van,andrei2020graphene,tran2020moire,regan2022emerging}. Similar to traditional semiconductor superlattices, these moir\'{e} superlattices  have large lattice constants ideal for observing Bloch oscillations and Wannier-Stark localization. However, the physics of these materials is much richer because they (1) often feature bands that carry nontrivial momentum-space quantum geometry, (2) offer {\it in-situ} band structure tunability via interlayer displacement field, substrate engineering, twist angle, and more, and (3) host a plethora of highly-correlated phases of matter. As such, moir\'e materials are attractive platforms for exploring novel physics that may occur at high electric fields. Recent theoretical analyses have already uncovered novel signatures in Wannier-Stark spectra and Bloch oscillating behavior of topological band structure models \cite{Krueckl2012Bloch,Gomez2013Floquet,long2017topological, Lee2015Direct,Kartashov2016Bloch,Topological2018Alexandradinata,Kolovsky2018Topological2, Kolovsky2018Topological, Li2019Bloch, Poddubny2019Distinguishing, Kim2020Anomalous,Li2020Tunable, di2020non,Zhu2021Uncovering, Liu2022Floquet,pan2023three} and of moir\'{e} materials \cite{Fahimniya2021Synchronizing, Vakhtel2022Bloch, Phong2023Quantum, de2023roses, De2023Berry, de2024floquet,zeng2024quantum}. In this work, we focus on the consequences of quantum geometry on Wannier-Stark spectra in two-dimensional band structures detectable through \textit{optical} techniques \cite{Voisin1984Optical,Bleuse1988Electric,Glutsch1998Excited}. Optical experiments provide a reliable means to identify the Wannier-Stark regime through direct detection of gaps in a ladder spectrum. In contrast, transport experiments are less definitive since they infer Bloch oscillations from negative differential conductance, which can often be explained by many different mechanisms of localization \cite{esaki1970superlattice,mendez1993wannier}. This inspires us to focus on optical absorption and emission. By adopting the Wannier-Stark basis, we accurately and efficiently capture the effects of large electric fields  for many-band systems in one and two dimensions. In particular, we re-cast the theory of Wannier-Stark localization in the modern language of non-Abelian Berry connection, and show how some of the geometric features can be ascertained in optical spectroscopy. Including interband processes is essential for moir\'{e} materials since their band structures are often highly-entangled. Finally, we propose specific experimental platforms on which these signatures should be accessible, highlighting the unique suitability of modern moir\'{e} materials.

\section{Quantum Geometric Theory}

We begin by recasting Wannier-Stark ladder theory in the language of non-Abelian Berry connection \cite{ adams1953crystal, kane1960zener, Carlo1994Theory, Glutsch1999Interaction, Lee2015Direct, Kolovsky2018Topological, Kolovsky2018Topological2, Poddubny2019Distinguishing, de2024floquet}.  We work with a tight-binding representation with $M$ orbitals per unit cell located at $\boldsymbol{\tau}_\sigma.$ Using the Fourier transform convention $\hat{c}^\dagger_{\mathbf{k}, \sigma} = N^{-\frac{1}{2}}\sum_{\mathbf{r}} e^{i \mathbf{k} \cdot \left( \mathbf{r} + \boldsymbol{\tau}_\sigma \right)}\hat{c}^\dagger_{\mathbf{r},\sigma},$ where $\mathbf{r}$ is a lattice translation vector and $N$ is the number of unit cells, the Hamiltonian takes the form \footnote{In principle, one can numerically solve Eq. \eqref{eq: hamiltonian 1} as an eigenvalue differential equation problem using a finite difference method. However, this system of differential equations is convection dominated. Therefore, in the process of obtaining the physical solutions, one needs to eliminate many unphysical solutions that are highly oscillatory. One reasonable approach is to insist on keeping only solutions whose eigenvalues are regularly spaced as the ladder structure dictates. However, even this restriction might not catch all pathological solutions. To be certain, one must  devise a method to inspect the eigenfunctions themselves to ensure that they are at least numerically differentiable. An alternative approach is to add a small amount of damping to the differential equations. We do not favor the finite difference approach in this work since it, though perfectly valid, obscures quantum geometric features encoded in the band structure.  }
\begin{equation}
\label{eq: hamiltonian 1}
    \hat{\mathcal{H}} = \sum_{\mathbf{k} \in \text{BZ},\sigma,\sigma'} \hat{c}^\dagger_{\mathbf{k},\sigma} \left[ \mathcal{H}_{0,\sigma,\sigma'}(\mathbf{k}) + i \delta_{\sigma,\sigma'} e \boldsymbol{\mathcal{E}}\cdot \nabla_\mathbf{k} \right] \hat{c}_{\mathbf{k},\sigma'},
\end{equation}
where $\boldsymbol{\mathcal{E}}$ is the electric field and $\mathcal{H}_0(\mathbf{k})$ is the $M\times M$ first-quantized Hamiltonian in the absence of the electric field. Eq. \eqref{eq: hamiltonian 1}  assumes that the position operator $\hat{\mathbf{r}}$ is diagonal in the orbital basis and that the $\mathbf{k}$-gradient acts on the right. The spectrum of this Hamiltonian comprises a ladder structure. Explicitly, the commutation relation $ \left[ \mathcal{H}(\mathbf{k}), L_+(\mathbf{r}) \right] = e \boldsymbol{\mathcal{E}} \cdot \mathbf{r} L_+(\mathbf{r}),$ where $L_+(\mathbf{r}) = e^{-i \mathbf{k} \cdot \mathbf{r}}\mathbb{1},$  implies that a state $\Omega_E$ at energy $E$ has a partner state at $E+  e\boldsymbol{\mathcal{E}} \cdot \mathbf{r}$ obtained by $\Omega_{E+  e\boldsymbol{\mathcal{E}} \cdot \mathbf{r}} = L_+(\mathbf{r}) \Omega_E.$ If $\mathbf{r}$ is perpendicular to $\boldsymbol{\mathcal{E}},$ states generated this way are degenerate (but not necessarily independent) because they are translated along equipotential lines. In fact, the Hamiltonian is translationally invariant along the direction perpendicular to $\boldsymbol{\mathcal{E}}.$ Therefore, we can retain the plane-wave character of energy eigenstates in that direction.

The Hamiltonian in Eq. \eqref{eq: hamiltonian 1} can be rewritten in the band basis wherein $\mathbf{k}$-space band geometry becomes apparent. For simplicity, we assume that there are no band degeneracies from $\mathcal{H}_0$ throughout this work \footnote{If degeneracies are present, then the formalism still works, but it should be applied directly to the orbital basis without first transforming into the band basis, which is useful when bands can be disentangled. In particular, we can write $\mathcal{H}_\text{L}(\mathbf{k}) = \mathcal{L}^\dagger(\mathbf{k}) \mathcal{H}(\mathbf{k})\mathcal{L}(\mathbf{k}),$ where $\mathcal{L}(\mathbf{k}) = \exp \left[\frac{i}{e\mathcal{E}} \int_0^{k_\parallel} \mathcal{H}_{0,\text{d}}(k_\parallel',k_\perp) dk_\parallel'\right].$ Then $\mathcal{H}_\text{L}(\mathbf{k}) = \mathcal{H}_{0,\text{nd}}(\mathbf{k}) + i \mathbb{1}e\mathcal{E} \partial_{k_\parallel}.$ This Hamiltonian is \textit{not} periodic in $\mathbf{k}$ for our choice of gauge. So care must be taken to choose suitable eigenfunctions that satisfy appropriate boundary conditions, as detailed in the text.}. Diagonalizing $\mathcal{H}_0(\mathbf{k}) = \mathcal{V}(\mathbf{k})\Lambda(\mathbf{k}) \mathcal{V}^\dagger(\mathbf{k}),$ where $\mathcal{V}(\mathbf{k})$ is the matrix of eigenvectors on its columns and $\Lambda(\mathbf{k})$ is the diagonal energy matrix, the Hamiltonian in the band representation now contains the term $ e\boldsymbol{\mathcal{E}} \cdot \boldsymbol{\mathcal{A}}(\mathbf{k})$ \cite{kane1961theory,Fukuyama1973Tightly,Carlo1994Theory}, where $\boldsymbol{\mathcal{A}}(\mathbf{k}) = i \mathcal{V}^\dagger(\mathbf{k}) \nabla_\mathbf{k}\mathcal{V}(\mathbf{k})$ is the non-Abelian Berry connection matrix. For this representation to be useful, a differentiable and periodic gauge for $\mathcal{V}(\mathbf{k})$ has to be chosen along the direction parallel to $\boldsymbol{\mathcal{E}},$ which can always be done even in topological bands \cite{Marzari1997Maximally,Marzari2012Maximally}. The Berry connection along the perpendicular direction never enters the Hamiltonian, so a differentiable gauge needs not be chosen there. The above considerations suggest that it is convenient to separate these two directions explicitly. We assume that $\boldsymbol{\mathcal{E}}$ is parallel to $\mathbf{b}_\parallel = n_1\mathbf{b}_1+ n_2 \mathbf{b}_2,$ where $n_1$ and $n_2$ are coprime integers and $\mathbf{b}_i$ are primitive reciprocal lattice vectors, and we  define $\mathbf{b}_\perp$ such that $\mathbf{b}_\parallel \cdot \mathbf{b}_\perp = 0$ and $|\mathbf{b}_\parallel \times \mathbf{b}_\perp | = |\mathbf{b}_1 \times \mathbf{b}_2|$ \footnote{Explicitly, we have $\mathbf{b}_\perp = \frac{\mathbf{b}_2 \cdot \mathbf{b}_\parallel}{\mathbf{b}_\parallel \cdot \mathbf{b}_\parallel} \mathbf{b}_1 - \frac{\mathbf{b}_1 \cdot \mathbf{b}_\parallel}{\mathbf{b}_\parallel \cdot \mathbf{b}_\parallel} \mathbf{b}_2$ for the perpendicular direction. The overall sign is not fixed}. The Hamiltonian is now partitioned into a \textit{matrix} differential equation along $k_\parallel$ for every value of $k_\perp,$ with $\mathcal{E} = |\boldsymbol{\mathcal{E}}|,$
\begin{equation}
\begin{split}
\label{eq: Hamiltonian 2}
    \mathcal{H}_\text{B}(k_\parallel,k_\perp) &= \Lambda(k_\parallel, k_\perp) + e \mathcal{E}  \mathcal{A}_\parallel(k_\parallel, k_\perp) + \mathbb{1} i e\mathcal{E} \partial_{k_\parallel},
    \end{split}
\end{equation}
where $ \mathcal{A}_\parallel = \boldsymbol{\mathcal{A}} \cdot \mathbf{b}_\parallel|\mathbf{b}_\parallel|^{-1}.$ For later convenience, we combine the diagonal terms into $\mathcal{D} = \Lambda + e \mathcal{E} \mathcal{A}_{\parallel,\text{d}}.$ Throughout, the subscripts ``$\text{d}$'' and ``$\text{nd}$'' to a matrix denote taking only that matrix's diagonal and non-diagonal elements respectively. Thus, the two-dimensional problem has been  reduced to a series of one-dimensional problems, one for each $k_\perp$ \cite{gluck2002wannier, hartmann2004dynamics, Thommen2002Theoretical, Yashima2003Dynamic, Hino2005Zener, Maksimov2015Wanniera, Maksimov2015Wannierb}.

Before proceeding further, let us consider the limit where all the off-diagonal couplings are zero, i.e. $\mathcal{A}_{\parallel,\text{nd}}(k_\parallel, k_\perp) = \mathbb{0}$. In this case, the solutions to Eq. \eqref{eq: Hamiltonian 2} can be written explicitly:
\begin{equation}
\label{eq: wave function in ladder basis}
\begin{split}
    \phi_{n,m}(k_\parallel, k_\perp) &=   e^{\frac{i}{e\mathcal{E}} \int_0^{k_\parallel} \left[\mathcal{D}(k_\parallel',  k_\perp) - E_{n,m}(k_\perp) \mathbb{1} \right] dk_\parallel'} \frac{\mathbb{1}_m}{\sqrt{N_\parallel}}\\
    E_{n,m}(k_\perp) &= \bar{\mathcal{D}}_{m,m}(k_\perp) + ne\mathcal{E} a_\parallel,
\end{split}
\end{equation}
where $a_\parallel = 2\pi |\mathbf{b}_\parallel|^{-1},$ $\mathbb{1}_m$ is the column vector of zeros everywhere except for one on the $m^\text{th}$ row, and $\bar{\mathcal{D}}_{m,m}(k_\perp) = \frac{a_\parallel}{2\pi} \int_0^{2\pi/a_\parallel} \mathcal{D}_{m,m}(k_\parallel, k_\perp) dk _\parallel$ is the average value of $\mathcal{D}_{m,m}(k_\parallel,k_\perp).$ We shall use overlines to denote the average of a function $f(k_\parallel, k_\perp)$ along the $k_\parallel$ direction throughout this work: $\bar{f}(k_\perp) = \frac{a_\parallel}{2\pi} \int_0^{2\pi/a_\parallel} f(k_\parallel,k_\perp) dk_\parallel$.  The spectrum contains a familiar ladder structure labeled by spatial index  $n$ and band index $m.$ In addition, there is dispersion in the $k_\perp$ direction due to both the band dispersion of the original Hamiltonian and the variation of the hybrid Wannier centers \cite{Taherinejad2014Wannier,Alexandradinata2014Wilson}, with the latter contribution being multiplied by $\mathcal{E}$. By hybrid Wannier functions, we mean those which are simultaneously defined in reciprocal space in one direction and in real space in the other. As such, the \textit{intra}band topology in the original band structure is encoded in $E_{n,m}(k_\perp).$ For bands whose Wannier centers do not wind along the $k_\perp$ direction, such as topologically-trivial bands, $E_{n,m}(k_\perp) = E_{n,m}(k_\perp + 2\pi/a_\perp),$ where $a_\perp = 2\pi |\mathbf{b}_\perp|^{-1}.$ For topological bands with a nontrivial winding, such as $\mathbb{Z}_2$ topological insulators or Chern insulators, we instead have $E_{n,m}(k_\perp) = E_{n,m}(k_\perp + 2\pi/a_\perp) + w e \mathcal{E} a_\parallel$ \cite{Lee2015Direct}, where $w$ is the winding number.

Even in topologically-trivial bands where the hybrid Wannier centers do not wind, they can still vary as a function of $k_\perp$. That is, even topologically-trivial bands can carry nontrivial band \textit{geometry} that can induce intersections of  Wannier-Stark bands in the limit of vanishing interband tunneling. These intersections always occur between different bands $(m\neq m')$ either with the same spatial index $(n=n')$ or with different spatial indices $(n \neq n')$. At these intersections, interband hybridization, i.e., Zener tunneling \cite{zener1934theory},  will generally gap them out. Thus, interband processes are crucial in multi-band systems, especially at large $\mathcal{E},$ and cannot be neglected or treated perturbatively in general. To analyze them, we perform a unitary transformation to absorb all \textit{intra}band effects and leave only \textit{inter}band processes for further scrutiny: $\mathcal{H}_\text{L}(\mathbf{k}) = \mathcal{L}^\dagger(\mathbf{k}) \mathcal{H}_\text{B}(\mathbf{k})\mathcal{L}(\mathbf{k}),$ where $\mathcal{L}(\mathbf{k}) = \exp \left[\frac{i}{e\mathcal{E}} \int_0^{k_\parallel} \mathcal{D}(k'_\parallel,k_\perp) dk_\parallel' \right].$ The Hamiltonian in this \textit{ladder representation} is just
\begin{equation}
\label{eq: Hamiltonian 3}
    \mathcal{H}_\text{L}(k_\parallel,k_\perp) =  e \mathcal{E}  \mathcal{A}_\text{L}(k_\parallel, k_\perp) + \mathbb{1} i e\mathcal{E} \partial_{k_\parallel},
\end{equation}
where $\mathcal{A}_\text{L}(\mathbf{k}) = \mathcal{L}^\dagger(\mathbf{k}) \mathcal{A}_{\parallel,\text{nd}}(\mathbf{k})\mathcal{L}(\mathbf{k})$ is purely off-diagonal. We can use the transformed ladder eigenfunctions $\psi_{n,m}(\mathbf{k}) = \mathcal{L}^\dagger(\mathbf{k}) \phi_{n,m}(\mathbf{k})$ as basis vectors \footnote{These basis states satisfy the following completeness relation $\sum_{k_\parallel} \psi_{n,m}^\dagger(\mathbf{k}) \psi_{n',m'}(\mathbf{k}) = \delta_{n,n'}\delta_{m,m'}$}. Here, they are simply plane-waves. We can, therefore, expand the Hamiltonian using the Fourier transform of $\mathcal{A}_\text{L}.$ Writing an eigenstate as $\Psi(k_\parallel,k_\perp) = \sum_{n,m} c_{n,m}(k_\perp)\psi_{n,m}(k_\parallel,k_\perp)$, the coefficients $c_{n,m}(k_\perp)$ are obtained from diagonalizing the  matrix equation
\begin{equation}
\label{eq: Hamiltonian 4}
\begin{split}
     E(k_\perp) c_{n,m}(k_\perp) = \left[\bar{\mathcal{D}}_{m,m}(k_\perp) + n e\mathcal{E}a_\parallel \right] c_{n,m}(k_\perp) \\
     + e \mathcal{E} \sum_{n',m'} \tilde{\mathcal{A}}_{\text{L},m,m'}(n-n',k_\perp) c_{n',m'}(k_\perp) ,\\
    \tilde{\mathcal{A}}_{\text{L},m,m'}(n,k_\perp) = \frac{a_\parallel}{2\pi}\int_0^{2\pi/a_\parallel} dk_\parallel \mathcal{A}_{\text{L},m,m'}(k_\parallel,k_\perp) \times \\
    \times e^{\frac{i}{e\mathcal{E}} \left[\bar{\mathcal{D}}_{m,m}(k_\perp) - \bar{\mathcal{D}}_{m',m'}(k_\perp)   \right]k_\parallel + ink_\parallel a_\parallel}.
\end{split}
\end{equation}
Although Eq. \eqref{eq: Hamiltonian 4} is formally a matrix equation of infinite size, we truncate it to a finite order in numerical diagonalization. The number of sites needed in a calculation depends on how fast the Fourier harmonics decay. We emphasize that this formalism is \textit{not} perturbative in $\mathcal{E},$ as is clear from factors $\mathcal{E}^{-1}$ in the exponential. In fact, for larger $\mathcal{E},$ we find that fewer Fourier harmonics of $\mathcal{A}_\text{L}$ are needed for an accurate calculation. As mentioned previously, we  find that ladder rungs belonging to the same band do not interact \textit{directly} because $\mathcal{A}_\text{L}$ is off-diagonal. Because of translational symmetry, we can still label states by a position index $n$ and a band index $m$. It is important to emphasize that while the existence of Wannier-Stark ladders can, in some cases, be explained purely by spectral considerations of the underlying band structure, the mixing between ladders and the resulting Zener-tunneled gaps are always inherently driven by interband Berry connection, as is made explicit in this formalism. 

\begin{figure*}
    \centering
    \includegraphics[width=6in]{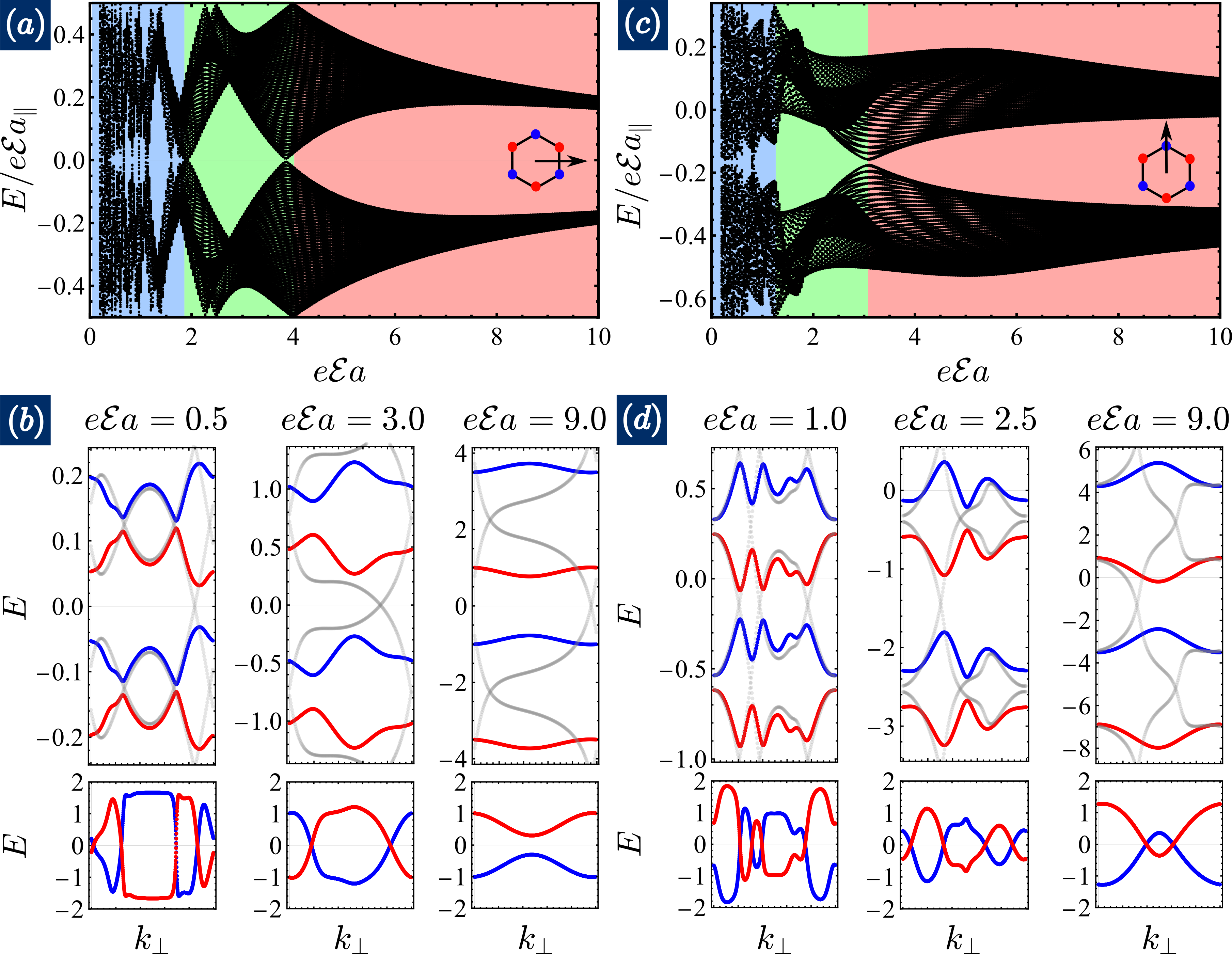}
    \caption{\textbf{Wannier-Stark spectra for the Haldane model as a function of field strength and direction.} The electric field points along $\hat{x}$ in (a,b) and along $\hat{y}$ in (c,d). In (a) and (c), the energies are plotted as a function of electric field. The background colors are used to qualitatively distinguish between the regimes of low (blue), moderate (green), and high (red) fields. A few representative band structures from each regime are shown in (b) and (d). For comparison, spectra \textit{without} interband tunneling are shown in light gray in the background. In the low-field regimes, we observe that the decoupled bands remain mostly intact except at places where those bands cross. In the large-field regimes, the bands become much flatter. Below each band structure, we also plot the renormalized energies which are obtained by removing the Stark contribution as detailed in the text. Energies are recorded in units of $t_1.$ Here, $a_\parallel=a/2$ in (a,b) and $a_\parallel=\sqrt{3}a/2$ in (c,d).}
    \label{fig:Haldane model}
\end{figure*}

At this point, it is worth specifying the boundary conditions of the wave functions in the various representations. In the orbital representation, $\mathcal{H}(\mathbf{k}) = \mathcal{H}_0(\mathbf{k}) +ie\boldsymbol{\mathcal{E}} \cdot \nabla_\mathbf{k}$ is \textit{not} periodic in $\mathbf{k},$ but instead obeys $\mathcal{H}(\mathbf{k}+\mathbf{G}) = \mathcal{B}(\mathbf{G})\mathcal{H}(\mathbf{k}) \mathcal{B}(\mathbf{G})^\dagger,$ where $\mathcal{B}_{\sigma,\sigma'}(\mathbf{G}) = \delta_{\sigma,\sigma'}e^{-i\mathbf{G} \cdot \boldsymbol{\tau}_\sigma}$ and $\mathbf{G}$ is a reciprocal lattice vector. Eigenfunctions, $\mathcal{H}(\mathbf{k}) \Omega(\mathbf{k}) = E\Omega(\mathbf{k}),$ satisfy $\Omega(\mathbf{k} + \mathbf{G}) = \mathcal{B}(\mathbf{G})\Omega(\mathbf{k}),$ and thus are not periodic. In the band representation, $\mathcal{H}_\text{B}(\mathbf{k}) = \mathcal{V}(\mathbf{k})^\dagger \mathcal{H}(\mathbf{k}) \mathcal{V}(\mathbf{k})$  and $\Phi(\mathbf{k}) = \mathcal{V}(\mathbf{k})^\dagger\Omega(\mathbf{k}).$ Using $\mathcal{V}(\mathbf{k}+\mathbf{G}) = \mathcal{B}(\mathbf{G})\mathcal{V}(\mathbf{k}),$ we find that $\Phi(\mathbf{k}+ \mathbf{G}) = \Phi(\mathbf{k})$ is periodic in $\mathbf{k}.$ Finally, in the ladder representation, $\mathcal{H}_\text{L}(\mathbf{k}) = \mathcal{L}(\mathbf{k})^\dagger \mathcal{H}_\text{B}(\mathbf{k})\mathcal{L}(\mathbf{k}),$ where $\mathcal{L}(\mathbf{k}) = \exp \left[\frac{i}{e\mathcal{E}} \int_0^{k_\parallel} \mathcal{D}(k'_\parallel,k_\perp) dk_\parallel' \right],$ and $\Psi(\mathbf{k}) = \mathcal{L}(\mathbf{k})^\dagger \Phi(\mathbf{k}).$ The boundary condition follows from $\mathcal{L}(k_\parallel+G_\parallel,k_\perp) =  \exp \left[\frac{i}{e\mathcal{E}} \bar{\mathcal{D}}(k_\perp) G_\parallel \right]\mathcal{L}(k_\parallel,k_\perp):$ $\Psi(k_\parallel+G_\parallel,k_\perp) =\exp \left[-\frac{i}{e\mathcal{E}} \bar{\mathcal{D}}(k_\perp) G_\parallel \right]\Psi(k_\parallel,k_\perp). $  It is clear that the basis functions $\psi_{n,m}(\mathbf{k}) = \mathbb{1}_mN_\parallel^{-1/2} \exp \left[-\frac{i}{e\mathcal{E}} E_{n,m}(k_\perp)k_\parallel \right]$ satisfy the correct boundary condition. So expanding using these basis functions is guaranteed to give the correct boundary condition.

The above analysis allows us to draw a qualitative distinction between the small-$\mathcal{E}$ and  large-$\mathcal{E}$ limits. In the former scenario, roughly when $e\mathcal{E}a_\parallel \ll \mathcal{W}$, where $\mathcal{W}$ is the bandwidth of the pristine band structure \footnote{The bandwidth is one possible measure of the dispersion of the bands. Since these bounds are only approximate, one can use other measures such as some well-defined matrix norm of $\mathcal{H}_0(\mathbf{k})$}, the decoupled energy bands capture the essence of the spectrum, including the topological winding of the bands. If the bands wind, they must intersect \footnote{Indeed, intersections must occur if the bands topologically wind across $k_\perp$, but these intersections need not occur at the same spatial index as shown in Fig. \ref{fig:schematic_figure} for emphasis. A simple example is when two bands have a large band gap.}; bands that do not wind can intersect too, but are not required to do so. At momenta $k_\perp^*$ where band intersections occur, gaps $\Delta \varepsilon$ form with magnitude  given by lowest-order degenerate perturbation theory:
\begin{equation}
\label{eq: gap estimation}
    \Delta \varepsilon \sim 2 e\mathcal{E} |\tilde{\mathcal{A}}_{\text{L},m,m'}(n,k_\perp^*)|,
\end{equation}
as shown in Fig. \ref{fig:schematic_figure}. We note that the gap is magnified by $\mathcal{E};$ so this approximation only holds for small $\mathcal{E}.$ Also, the order of a gap $n$ denotes the hybridization between Wannier-Stark states that are separated by $|n|$ lattice sites apart. Thus, the gaps generally decrease with increasing $|n|.$ In the high-$\mathcal{E}$ limit, the Wannier-Stark bands no longer resemble the decoupled bands since interband processes dominate the spectral formation. In fact, in the $e\mathcal{E}a_\parallel \gg \mathcal{W}$ limit, the band basis adopted here is not necessarily the most physically transparent. Instead, one should remain in the orbital basis in Eq. \eqref{eq: hamiltonian 1} and treat $\mathcal{H}_{0,\sigma,\sigma'}(\mathbf{k})$ as a perturbation on $i \delta_{\sigma,\sigma'} e \boldsymbol{\mathcal{E}}\cdot \nabla_\mathbf{k}$. In this case, the Wannier-Stark spectrum consists essentially of the Stark energy on localized orbitals, which is dispersionless in $k_\perp,$ as shown in Fig. \ref{fig:schematic_figure}. Any sense of band topology that is encoded in the entanglement between orbitals becomes difficult to discern. We do not consider the strict $\mathcal{E} \rightarrow \infty$ regime in this work, for it is difficult to access experimentally. Instead, any mention of ``high field" here refers to parameter regimes where low-order perturbation theory cannot capture interband tunneling. Gaps induced by interband hybridizations have recently also been derived in Ref. \cite{de2024floquet} using a time-dependent Floquet formalism.

\section{Prototypical Model}

For illustration, we now apply this formalism to a prototypical model of a topological insulator: the Haldane model \cite{Haldane1988Model}. In the orbital basis, the Hamiltonian is $\mathcal{H}_0(\mathbf{k}) = \mathbf{h}(\mathbf{k}) \cdot \boldsymbol{\sigma},$ where $\sigma_i$ are Pauli matrices, $h_x(\mathbf{k})-ih_y(\mathbf{k}) = -t_1\sum_{\boldsymbol{\delta}_i}e^{i \mathbf{k} \cdot \boldsymbol{\delta}_i},$ $h_z = -t_2 \sum_{\mathbf{a}_i} \sin\left(\mathbf{k}\cdot \mathbf{a}_i \right).$ This model has only one adjustable parameter $t_2/t_1 \neq 0$ that simultaneously controls the gap size, the bandwidth, and the topology of the bands; in practice, $|t_2/t_1| \ll 1,$ so we fix $t_2/t_1 = 0.1$  for simplicity. The topology of the Haldane model is discussed in detail in Appendix \ref{sec: Details on the Haldane Model}. In this Appendix, we also show how to choose a continuous gauge of the wave functions for the Haldane model, using the method shown in Refs. \cite{Marzari1997Maximally,Marzari2012Maximally}.  First, we apply the electric field in the $x$-direction. As shown in Fig. \ref{fig:Haldane model}(a) in the blue and green regions, various global gaps develop at small electric fields that evolve non-monotonically as field strength enhances. There are certain values of $\mathcal{E}$ at which the global gaps are quenched entirely. However, these precise gap closures require fine tuning of parameters, so we shall not devote particular attention to them in this work. The convoluted evolution of the band structure at small $\mathcal{E}$ values reflects the fact there are a variety of competing processes operative in this regime, namely effects from both spectral and quantum geometrical interband and intraband properties. When $\mathcal{E}$ becomes large, the physics is universally dominated by on-site Stark energy. Therefore, the bands become monotonically flatter as electric field increases, as shown in the red region of Fig. \ref{fig:Haldane model}(a). For particular values of $\mathcal{E},$ we show the momentum-resolved band structures in Fig. \ref{fig:Haldane model}(b). For small $\mathcal{E},$ the band structures somewhat resemble their counterparts \textit{without} Zener tunneling, which are shown in gray in Fig. \ref{fig:Haldane model}(b). Interband hybridization results in gaps that can be estimated by Eq. \eqref{eq: gap estimation}.  Such a resemblance is lost when $\mathcal{E}$ is large. The physics is qualitatively the same when $\mathcal{E}$ points along the $y$-direction, as shown in Fig. \ref{fig:Haldane model}(c,d).

\begin{figure*}
    \centering
    \includegraphics[width=5in]{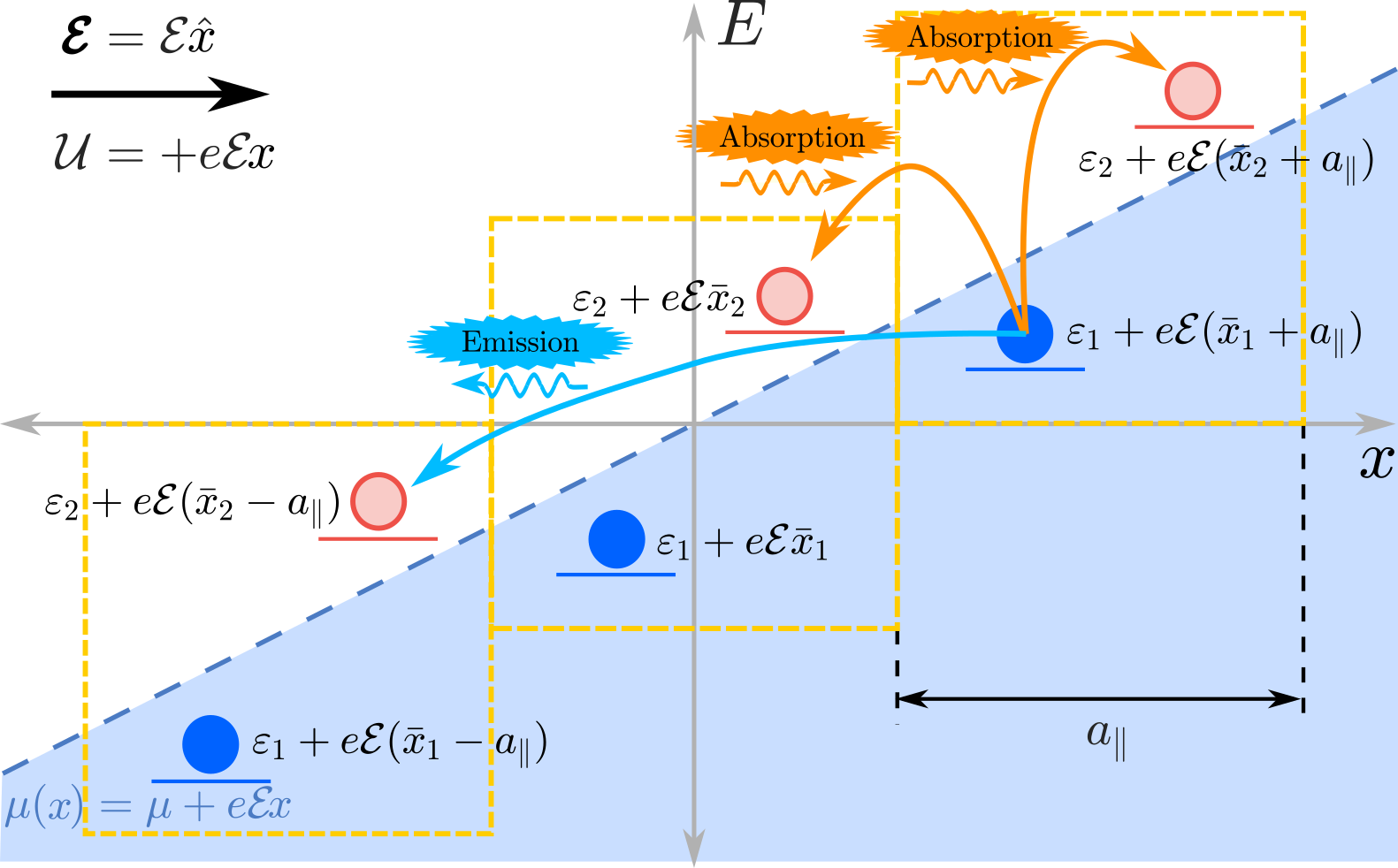}
    \caption{\textbf{Schematic diagram of electron occupation in one dimension.} The  dashed blue line is the spatially-dependent chemical potential $\mu(x).$ States below $\mu(x)$ are filled (blue circles) and above are unfilled (red). Transitions from a filled lower (higher) energy state to an unfilled higher (lower) energy state correspond to an absorption (emission) process. Yellow boxes define the unit cells. }
    \label{fig:chemical_potential.png}
\end{figure*}

\section{Optical Properties}

\begin{figure*}
    \centering
    \includegraphics[width=6in]{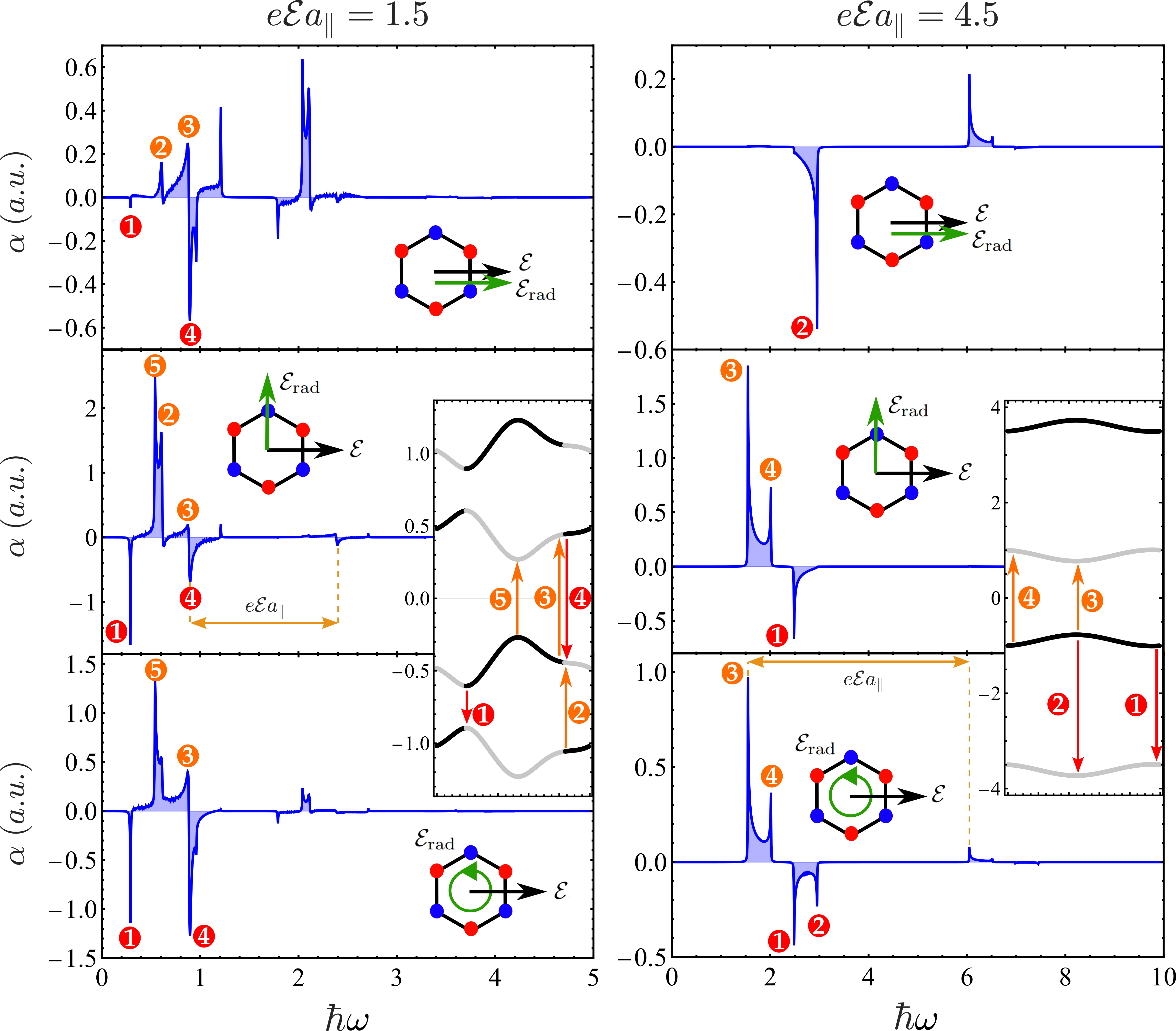}
    \caption{\textbf{Optical spectra for the Haldane model.} The \textit{net} absorption $\alpha$ is plotted for $e\mathcal{E}a_\parallel = 1.5$ and $4.5$ where the radiation field is linearly polarized parallel to the static field (top panels), is linearly polarized perpendicular to the static field (middle panels), or is circularly polarized (bottom panels). For both moderate and high electric fields, either absorption or emission dominates depending on input frequency. A few prominent peaks and their corresponding transitions are numerically indicated. Many less pronounced peaks can be seen; their transition pathways can be mapped to the band structure as well. To avoid clutter, we leave these unlabeled. In the inset band structures, occupied states are black while unoccupied states are  light gray. A few (not all) replica peaks separated by $e\mathcal{E}a_\parallel$ are indicated with orange arrows. Here, the mesh sizes along the parallel and perpendicular directions are $N_\parallel = 200$ and $N_\perp = 350$ respectively, and the broadening factor of the Dirac delta function is $\eta = 0.003$.  }
    \label{fig:optical spectra}
\end{figure*}

The interband-hybridized gaps can, in principle, be probed by THz  spectroscopy. In order to calculate various optical responses, it is necessary to determine electron occupation. Generally, this is a highly nontrivial, non-equilibrium problem requiring assumption about various relaxation pathways \cite{Wacker1998Quantum,mahan2013many,stefanucci2013nonequilibrium}. In this work, we prefer to focus on \textit{qualitative} features of possible optical signatures and delegate detailed investigations to future studies. To this end, we  assume the simplest physically-motivated form of the occupation function based on translational symmetry. That is, we \textit{assume} the electron density is uniform on the unit-cell scale. This  is also consistent with the existence of local charge neutrality. Inhomogeneous charge distributions are likely to be compensated by intra-unit-cell charge transfers. This effect is captured by the screening effect described in the Hartree approximation (it is worth noting that this can be the case in magic angle twisted bilayer graphene). Then, the occupation of states shifted over by integer multiples of $a_\parallel$ must be the same. From these considerations, we are  led to a spatially-varying chemical potential, which is permitted since we are not in equilibrium: $\mu_{n,m}(k_\perp) = \mu + e \mathcal{E}\bar{\mathbf{r}}_{\parallel,n,m}(k_\perp),$ where $\bar{\mathbf{r}}_{\parallel,n,m}(k_\perp)$ is the average position of the state being evaluated. Similar forms of the occupation function were used in previous studies  \cite{Tsu1975Hopping,dohler1975new,Shon1996Hopping, rott1997hopping,Wacker1998Quantum, rott1999self, wacker2002semiconductor}. As a matter of convenience, we continue to use the Fermi-Dirac distribution:
\begin{equation}
    f\left[E_{n,m}(k_\perp)  \right]  = \left[1+e^{\beta\left[E_{n,m}(k_\perp)-\mu - e \mathcal{E}\bar{\mathbf{r}}_{\parallel,n,m}(k_\perp)\right] }\right]^{-1},
\end{equation}
where $\beta\rightarrow +\infty$ is inverse temperature. Despite appearance, one should \textit{not} regard the utilization of the Fermi-Dirac distribution as implying an equilibrium situation. This assignment of electron occupation is sensible in both weak and large electric field limits. For weak fields, $\mathcal{E} \rightarrow 0,$ where the chemical potential approaches homogeneity, the distribution function reverts to the equilibrium Fermi-Dirac distribution. In the large-field limit, and as $\mathcal{E} \rightarrow \infty,$ the occupation is that of basically dispersionless Wannier-Stark bands. Another consistency check is to consider the limit where all interband couplings are switched off. Then, the total energies are given analytically by Eq. \eqref{eq: wave function in ladder basis}. The assumed chemical potential prescribes that only the \textit{original} band energies $\Lambda$ dictate the occupation of states and not the Stark contribution $e\mathcal{E}(\mathcal{A}_\parallel + na_\parallel),$ i.e. states that are occupied before $\mathcal{E}$ is turned on remain occupied after $\mathcal{E}$ is switched on. This is sensible for a gapped insulator (as in the Haldane model) where all the states of the valence bands are occupied both before and after $\mathcal{E}$ is introduced. In the absence of interband tunneling, such insulators must remain insulating no matter the strength of the electric field. We illustrate this occupation function schematically in Fig. \ref{fig:chemical_potential.png} for one dimension. In the two-dimensional case, in each $k_\perp$ sector, we essentially have a one-dimensional problem. Reassured by the preceding considerations, we expect this occupation function to be especially well-suited for charge-neutral systems with fully-filled bands \footnote{For partially-filled bands, the shift in Fermi surface depends on scattering time and is not captured in our approximation. This can be seen even in semi-classical Boltzmann transport}.  We point out that the combination $E- e \mathcal{E}\bar{\mathbf{r}}_{\parallel}$ is just the energy without the Stark energy contribution $e \mathcal{E}\bar{\mathbf{r}}_{\parallel}$. For the Haldane model, we plot this combination at the bottom of Fig. \ref{fig:Haldane model}(b,d).  Further discussion of the occupation function appears in Appendix \ref{sec: Assumed Occupation Function}.

Using Fermi's golden rule, we now calculate optical absorption and emission at charge neutrality. Under exposure to weak monochromatic radiation $\boldsymbol{\mathcal{E}}_\text{rad}$ with frequency $\omega>0,$ the stimulated absorption and emission coefficients are \cite{chuang2012physics}
\begin{equation}
\label{eq: optical formulas}
\begin{split}
    \mathfrak{a} (\omega) &\propto \omega \sum_{a,b}  \abs{\bra{b} \hat{d} \ket{a}}^2 \delta \left( E_b-E_a-\hbar \omega \right) f_a\left[1-f_b \right] ,\\
    \mathfrak{e} (\omega) &\propto \omega \sum_{a,b}\abs{\bra{b} \hat{d} \ket{a}}^2  \delta \left( E_a-E_b+\hbar \omega \right) f_b \left[ 1-f_a\right] ,    
\end{split}
\end{equation}
where $a$ and $b$ are generic labels for states.  Eq. \eqref{eq: optical formulas} is derived in Appendix \ref{sec: Optical Absorption}.  The net absorption coefficient is just the difference between the two: $\alpha(\omega) = \mathfrak{a} (\omega) - \mathfrak{e} (\omega).$ In practice, it is often $\alpha$ that is measured since disentangling between pure emission and absorption can be difficult. In equilibrium,  $\alpha \geq 0,$ that is, absorption dominates. However, when a system is pushed away from equilibrium, $\alpha <0$ is possible; this is the regime where the electronic platform amplifies the optical field. Here, $\hat{d} = e \hat{\mathbf{r}}_\text{rad}$ is the dipole operator projected into the radiation's oscillation direction. Taking advantage of the two-dimensional nature of the problem, we have the freedom to choose the direction and polarization of $\hat{d}$ independently of the static electric field. There is a subtlety in evaluating the expectation values of $\hat{d}$ when the states involved are extended. When $\hat{\mathbf{r}}_\text{rad} = \hat{\mathbf{r}}_\parallel,$ we simply have 
\begin{equation}
\label{eq: position operator}
\begin{split}
    &\bra{n',m',k_\perp} \hat{\mathbf{r}}_\parallel \ket{n,m,k_\perp} \\
    &=   i \sum_{k_\parallel} \Omega_{n',m'}^\dagger(k_\parallel,k_\perp) \frac{\partial}{\partial k_\parallel}  \Omega_{n,m}(k_\parallel,k_\perp) .
\end{split}
\end{equation} This is because along the parallel direction to the applied static field, states are localized. Thus, the matrix elements of the position operator are well-defined, even on the diagonal.  However, when $\hat{\mathbf{r}}_\text{rad} = \hat{\mathbf{r}}_\perp,$ the analogous procedure fails because the perpendicular derivative of wave functions cannot always be defined. We, therefore, adopt an alternative method \cite{vanderbilt2018berry}:
\begin{equation}
\label{eq: perpendicular r}
    \begin{split}
    &\bra{n',m',k_\perp}\hat{\mathbf{r}}_\perp\ket{n,m,k_\perp} \left[ E_{n,m}(k_\perp)-E_{n',m'}(k_\perp)\right] \\
    &= i\sum_{k_\parallel} \Omega_{n',m'}^\dagger(k_\parallel, k_\perp)  \frac{\partial \mathcal{H}_{0}(k_\parallel, k_\perp)}{\partial k_\perp}  \Omega_{n,m}(k_\parallel, k_\perp).
    \end{split}
\end{equation} Clearly, this formula only works for non-degenerate states, but that is all we need since degenerate states are eliminated from the sum by the occupation functions. Therefore, Eq. \eqref{eq: perpendicular r} has all the generality we require. For circularly polarized light, we choose $\hat{\mathbf{r}}_\text{rad} = 2^{-1/2} \left( \hat{\mathbf{r}}_\parallel \pm i \hat{\mathbf{r}}_\perp \right)$. Derivations of Eqs. \eqref{eq: position operator} and \eqref{eq: perpendicular r} are done in Appendix \ref{sec: Evaluation of Various Position Expectation Values}.

For the Haldane model, we show optical absorption results for some representative values of the applied fields in Fig. \ref{fig:optical spectra}. For each field strength $e\mathcal{E}a_\parallel = 1.5$ or $ 4.5,$ we plot the \textit{net} absorption spectra for linearly-polarized radiation parallel to (top panels) and perpendicular to (middle panels) the static $\boldsymbol{\mathcal{E}}$ field and for circularly-polarized radiation (bottom panels). A few prominent peaks are labeled, and their corresponding transitions in the energy spectra are indicated. Compared to the optical spectra of Wannier-Stark ladders in one-dimensional semiconductor superlattices, our spectra are considerably richer because of increased dimensionality and prominent Zener tunneling. Peaks that are replicated at frequency distances of integer multiples of the Bloch frequency $e\mathcal{E}a_\parallel,$  as shown in Fig. \ref{fig:optical spectra}, are transitions of the same pair of bands and momentum but differ by the real-space separation of the states, i.e., same $m,m',k_\perp$ but different $n,n'$. Not all peaks have prominent replicas because some matrix elements decay quickly with increasing rung separation. Just as in semiconductor superlattices, the existence of equally-spaced peaks with separation corresponding to integer multiples of the Bloch frequency can be taken as a smoking-gun signature of the Wannier-Stark regime.

In the large-field regime, we find that either absorption dominates or emission dominates depending on input frequency. Interestingly, in moderate fields, both absorption-dominated and emission-dominated regimes are also observed. Taken at its face value, this result implies that the lasing regime is achievable at reasonable electric fields in moir\'{e} materials---a claim to be confirmed with further microscopic modeling. The emission-dominated regimes are due to transitions from \textit{high-energy occupied} states to \textit{low-energy unoccupied} states where the atypical spectral alignment is maintained by the static electric field, as shown in Fig. \ref{fig:chemical_potential.png}. This is similar to the operation of a quantum cascade laser \cite{kazarinov1971possibility,faist1994quantum}. However, we emphasize a distinction: in moir\'{e} materials, the absorption-dominated frequency regime arises from interband hybridization (i.e., Zener-tunneled gaps) as well as from inter-ladder-site hopping (i.e., the dominant mechanism in semiconductor superlattices).  Furthermore, we find that peaks generically occur at frequencies corresponding to transitions between band maxima and minima. Thus, the most pronounced peaks inform us about the magnitude of gaps.

\section{Discussion}
    
The Haldane model is an appropriate low-energy effective model for several moir\'{e} platforms, such as twisted transition metal dichalcogenide homobilayers \cite{Wu2019Topological,Repellin2020Chern,devakul2021magic}. In these systems, the bandwidth is on the order of $10-100$ meV, and the superlattice constant is about $10-20$ nm. Consequently, the electric field scale $\mathcal{E} \sim t_0/ea$ needed to induce appreciable Zener-tunneling is about $5-20$ kV/cm, well within experimental capacity. The resulting interband-hybridized gaps are on the meV scale. They can be probed using THz spectroscopy. Therefore, we present moir\'{e} materials as attractive, realistic platforms to study optical absorption and emission induced by applying a \textit{static} electric field of moderate to high intensity. Such a field causes an otherwise continuous energy spectrum to form discrete Wannier-Stark bands that encode both intraband quantum geometry in the form of a band-projected winding number, and interband quantum geometry manifested as Zener-tunneled gaps. The possibility of stimulated emission dominating absorption for various parameter regimes raised by our investigation is exciting since it may have important implications for laser technologies. However, we caution that our conclusions are based on a simplified (but physically motivated \cite{Tsu1975Hopping,dohler1975new,Shon1996Hopping, rott1997hopping,Wacker1998Quantum, rott1999self, wacker2002semiconductor}) assumption of the occupation of electrons that should be verified using more sophisticated methods in future work. Though we expect our qualitative conclusion on the possibility of lasing to hold, those future investigations may lead to other insights presently overlooked. Furthermore, Coulomb interactions may lead to qualitatively new physics in the Wannier-Stark regime that warrants further scrutiny. For instance, there may be symmetry-broken states that can be stabilized by a non-negligible static electric field and probed by optical spectroscopy.

We thank Christophe De Beule, Roshan Krishna Kumar, Riccardo Bertini, and Frank Koppens for insightful discussions. V.T.P.  further acknowledges Christophe De Beule and Eugene J. Mele for previous related collaborations on Wannier-Stark localization and Bloch oscillations. V.T.P. and C.L. are supported by start-up funds from Florida State University and the National High Magnetic Field Laboratory. The National High Magnetic Field Laboratory is supported by the National Science Foundation through NSF/DMR-2128556 and the State of Florida. F.G. acknowledges support from the Severo Ochoa programme for centres of excellence in R\&D (CEX2020-001039-S / AEI / 10.13039/501100011033, Ministerio de Ciencia e Innovaci\'on, Spain);  from the grant (MAD2D-CM)-MRR MATERIALES AVANZADOS-IMDEA-NC, NOVMOMAT, Grant PID2022-142162NB-I00 funded by MCIN/AEI/ 10.13039/501100011033 and by “ERDF A way of making Europe”.

\appendix

\onecolumngrid

\section{Details on the Haldane Model}
\label{sec: Details on the Haldane Model}

The Haldane model is a prototypical model of a topological insulator. Thus, it serves as an excellent example for the examination of topological signatures in Wannier-Stark spectra. The primitive lattice vectors are 
\begin{equation}
    \mathbf{a}_1 = a \left( 1, 0 \right) \quad \text{and} \quad \mathbf{a}_2 = a \left( -\frac{1}{2}, \frac{\sqrt{3}}{2} \right).
\end{equation}
We also define $\mathbf{a}_3 = -\mathbf{a}_1-\mathbf{a}_2 .$ The primitive reciprocal lattice vectors are 
\begin{equation}
    \mathbf{b}_1 = \frac{4\pi}{\sqrt{3}a} \left(\frac{\sqrt{3}}{2},\frac{1}{2} \right)  \quad \text{and} \quad \mathbf{b}_2 = \frac{4\pi}{\sqrt{3}a} \left(0,1 \right).
\end{equation}
We clearly have $\mathbf{a}_i \cdot \mathbf{b}_j = 2\pi \delta_{ij}.$ We furthermore define nearest-neighbor vectors 
\begin{equation}
    \boldsymbol{\tau}_1 = \frac{a}{\sqrt{3}} \left( 0, 1 \right), \quad \boldsymbol{\tau}_2 = \frac{a}{\sqrt{3}} \left( -\frac{\sqrt{3}}{2}, -\frac{1}{2} \right), \quad \text{and} \quad \boldsymbol{\tau}_3 = \frac{a}{\sqrt{3}} \left( +\frac{\sqrt{3}}{2}, -\frac{1}{2} \right).
\end{equation}
Within a unit cell, we place the $A$ sublattice at $(0,0)$ and the $B$ sublattice at $\boldsymbol{\tau}_1.$ The Hamiltonian without an electric field is given by 
\begin{equation}
\begin{split}
        \mathcal{H}_0(\mathbf{k}) &=   \left[-t_1 \sum_{i=1}^3 \cos \left( \mathbf{k} \cdot \boldsymbol{\tau}_i \right) \right]\sigma_x+\left[t_1 \sum_{i=1}^3 \sin \left( \mathbf{k} \cdot \boldsymbol{\tau}_i \right) \right]\sigma_y + \left[-t_2 \sum_{i=1}^3 \sin \left( \mathbf{k} \cdot \mathbf{a}_i \right) \right]\sigma_z \\
        &= h_x(\mathbf{k})\sigma_x + h_y(\mathbf{k})\sigma_y  + h_z(\mathbf{k})\sigma_z.
\end{split}
\end{equation}
Using $\mathcal{B}(\mathbf{b}_1) = \begin{pmatrix}
    1 & 0 \\
    0 & e^{ - \frac{2\pi i }{3}}
\end{pmatrix}$ and $ \mathcal{B}(\mathbf{b}_2) = \begin{pmatrix}
    1 & 0 \\
    0 & e^{ + \frac{2\pi i }{3}}
\end{pmatrix},$ it is straightforward to verify that $\mathcal{H}_0\left(\mathbf{k} + \mathbf{b}_i \right) = \mathcal{B}(\mathbf{b}_i) \mathcal{H}_0(\mathbf{k}) \mathcal{B}(\mathbf{b}_i)^\dagger.$ Under the assumption that the system is gapped for all $\mathbf{k}$, we can write  $\mathbf{h}(\mathbf{k}) = \left(h_x(\mathbf{k}), h_y(\mathbf{k}), h_z(\mathbf{k})  \right),$ $\lambda(\mathbf{k}) = \abs{\mathbf{h}(\mathbf{k})},$ $\hat{h}(\mathbf{k}) = \mathbf{h}(\mathbf{k})/\abs{\mathbf{h}(\mathbf{k})},$ and $\mathcal{H}_0(\mathbf{k}) = \lambda(\mathbf{k}) \hat{h}(\mathbf{k}) \cdot \boldsymbol{\sigma}.$ $\lambda(\mathbf{k})$ is precisely the magnitude of the eigenenergies, so $\lambda(\mathbf{k})$ must be periodic in $\mathbf{k}.$ The transformation property of $\hat{h}(\mathbf{k})$ follows that of $\mathcal{H}_0(\mathbf{k}):$ $\hat{h}(\mathbf{k}+ \mathbf{b}_i) \cdot \boldsymbol{\sigma} = \mathcal{B}(\mathbf{b}_i) \left[\hat{h}(\mathbf{k}) \cdot \boldsymbol{\sigma} \right] \mathcal{B}(\mathbf{b}_i)^\dagger.$ Explicitly, we have the following 
\begin{equation}
    \begin{split}
        \hat{h}_z(\mathbf{k}+\mathbf{b}_i) &= \hat{h}_z(\mathbf{k}), \\
        \hat{h}_x(\mathbf{k}+\mathbf{b}_i)+i\hat{h}_y(\mathbf{k}+\mathbf{b}_i) &= \left[ \hat{h}_x(\mathbf{k})+i\hat{h}_y(\mathbf{k})\right] e^{-i \mathbf{b}_i \cdot \boldsymbol{\tau}_1}.
    \end{split}
\end{equation}

One possible periodic gauge for the eigenvectors is 
\begin{equation}
    v^\text{(I)}_\pm(\mathbf{k}) = \frac{1}{\sqrt{2\pm 2\hat{h}_z(\mathbf{k})}}\begin{pmatrix}
        \hat{h}_z(\mathbf{k})\pm 1\\
        \hat{h}_x(\mathbf{k})+i \hat{h}_y(\mathbf{k})
    \end{pmatrix}, \quad \lambda_\pm(\mathbf{k}) = \pm \lambda(\mathbf{k}).
\end{equation}
One notices that this gauge is numerically obtained by simply fixing the upper component of the eigenvectors to be real. This gauge is smooth everywhere except at isolated values of $\mathbf{k}$ for which $\hat{h}_z(\mathbf{k}) = \pm 1 .$ This can only occur when $\hat{h}_x(\mathbf{k}) = \hat{h}_y(\mathbf{k}) = 0,$ so $\mathbf{k} = \mathbf{K}_\pm = \frac{4\pi}{3a} \left( \pm 1, 0 \right)$ at the Dirac points (and its equivalents) in this specific model. At these points, $\hat{h}_z(\mathbf{K}_\pm) = \pm \text{sgn}(t_2).$ These singularities need to be patched before we can take derivatives. To do so, we use another possible periodic gauge for the eigenvectors 
\begin{equation}
    v^\text{(II)}_\pm(\mathbf{k}) = \frac{e^{-i\mathbf{k} \cdot \boldsymbol{\tau}_1}}{\sqrt{2\mp 2\hat{h}_z(\mathbf{k})}} \begin{pmatrix}
        \hat{h}_x(\mathbf{k}) - i \hat{h}_y(\mathbf{k}) \\
        \pm 1 - \hat{h}_z(\mathbf{k})
    \end{pmatrix}, \quad \lambda_\pm(\mathbf{k}) = \pm \lambda(\mathbf{k}).
\end{equation}
Numerically, this gauge is obtained by fixing the lower component of the eigenvectors to be real modulo an overall exponential factor of $e^{-i\mathbf{k} \cdot \boldsymbol{\tau}_1}$. This gauge is also smooth everywhere except at values of $\mathbf{k}$ for which $\hat{h}_z(\mathbf{k}) = \pm 1,$ similar to the previous gauge. However, the locations of the singularity in each band has been switched from $\mathbf{K}_\pm$ to $\mathbf{K}_\mp,$ as shown in Table \ref{Tab: Location of singularity}. As long as $k_\parallel$ does not traverse \textit{both} $\mathbf{K}_+$ and $\mathbf{K}_-,$ this allows us to choose either gauge I or gauge II to avoid the singularity depending of $k_\perp.$ For example,  let us consider  $\boldsymbol{\mathcal{E}} = \mathcal{E} \hat{y}.$ Then, we have $\mathbf{b}_\parallel = \mathbf{b}_2,$ $\mathbf{b}_\perp = \mathbf{b}_1- \frac{1}{2}\mathbf{b}_2 = \left(2\pi/a,0\right),$ $a_\parallel = \sqrt{3}a/2,$ and $a_\perp = a.$ Our choice of Brillouin zone is $k_\parallel \times k_\perp \in [0,4\pi/\sqrt{3}a) \times [0,2\pi/a).$ To avoid the singularities, we choose for $t_2 > 0$
\begin{equation}
    \mathcal{V}(\mathbf{k}) = \begin{cases}
    \begin{pmatrix}
        v_-^\text{(I)}(\mathbf{k}) & v_+^\text{(II)}(\mathbf{k})
    \end{pmatrix}, & \text{for } k_x \in [0, \pi/a) \\
     \begin{pmatrix}
        v_-^\text{(II)}(\mathbf{k}) & v_+^\text{(I)}(\mathbf{k})
    \end{pmatrix}, & \text{for } k_x \in [\pi/a, 2\pi/a)
  \end{cases},
\end{equation}
and for $t_2 < 0$
\begin{equation}
    \mathcal{V}(\mathbf{k}) = \begin{cases}
    \begin{pmatrix}
        v_-^\text{(II)}(\mathbf{k}) & v_+^\text{(I)}(\mathbf{k})
    \end{pmatrix}, & \text{for } k_x \in [0, \pi/a) \\
     \begin{pmatrix}
        v_-^\text{(I)}(\mathbf{k}) & v_+^\text{(II)}(\mathbf{k})
    \end{pmatrix}, & \text{for } k_x \in [\pi/a, 2\pi/a)
  \end{cases}.
\end{equation}

\begin{table}
\caption{\textbf{Location of singularity}.}
\begin{center}
\begin{tabular}{ |c||c||c| } 
\hline\hline
Singularity of ...  &  $t_2>0$ & $t_2<0$ \\
\hline\hline
 $v_-^\text{(I)}$ & $\mathbf{K}_+$ & $\mathbf{K}_-$ \\ 
 $v_+^\text{(I)}$ & $\mathbf{K}_-$ & $\mathbf{K}_+$ \\ 
 $v_-^\text{(II)}$ & $\mathbf{K}_-$ & $\mathbf{K}_+$ \\ 
 $v_+^\text{(II)}$ & $\mathbf{K}_+$ & $\mathbf{K}_-$ \\ 
\hline\hline
\end{tabular}
\label{Tab: Location of singularity}
\end{center}
\end{table}

\begin{figure}
    \centering
    \includegraphics[width=5.5in]{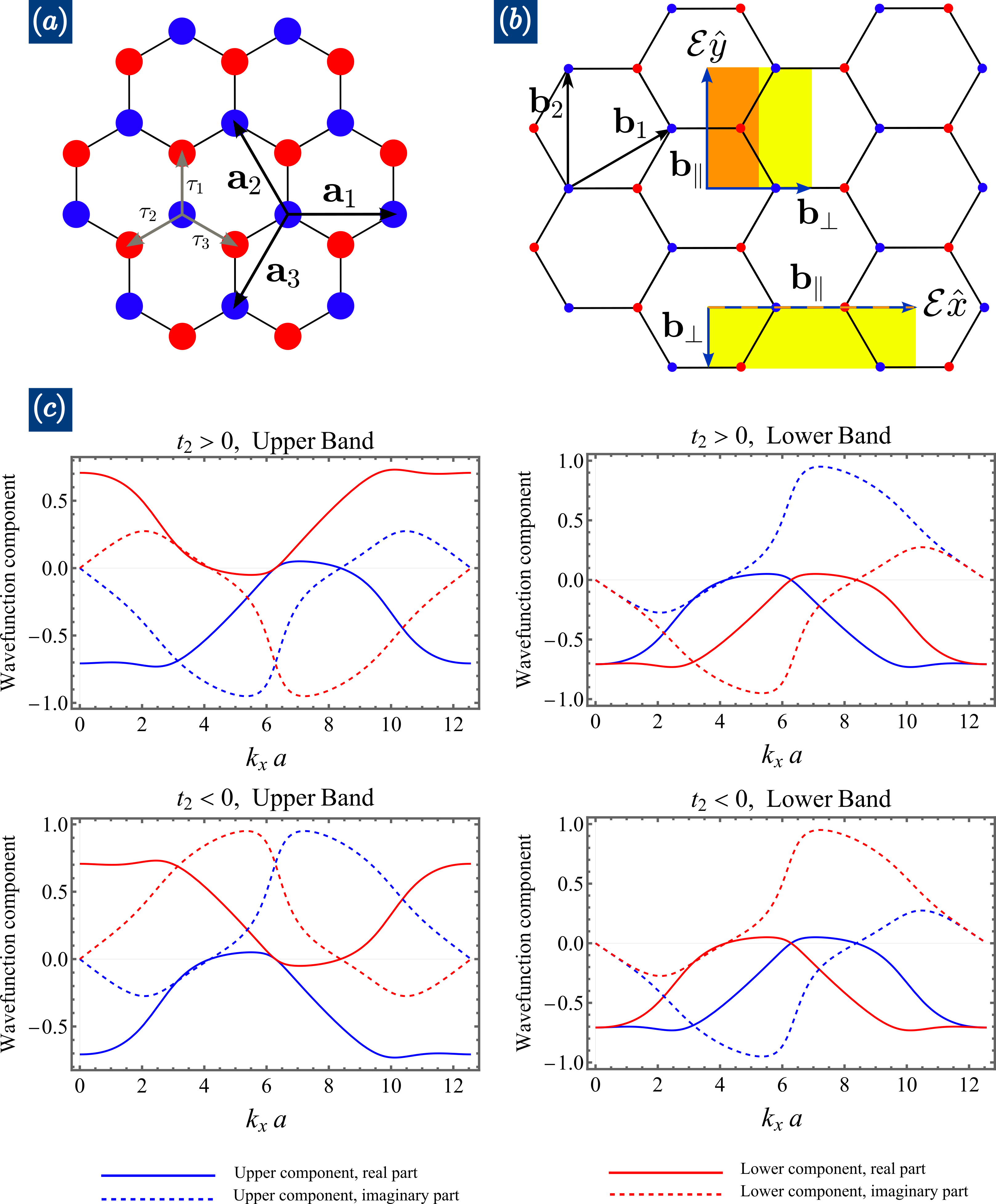}
    \caption{\textbf{Gauge fixing in the Haldane model.  }(a) Real-space representation of a hexagonal bipartite lattice with lattice vectors and nearest-neighbor vectors indicated. (b) Reciprocal-space representation of the Haldane model. Example Brillouin zones for when $\boldsymbol{\mathcal{E}} = \mathcal{E} \hat{y}$ and $\boldsymbol{\mathcal{E}} = \mathcal{E} \hat{x}$ are shown. In the former case, the Brillouin zone is shaded in two different colors, orange and yellow, to suggest that those regions should have  different gauges to avoid a singularity at  $\mathbf{K}_\pm.$ In the latter case, a single quasi-smooth gauge can be chosen everywhere in the Brillouin zone except along the line that runs through both $\mathbf{K}_\pm.$ There, care must be taken to choose a differentiable gauge along the direction $\mathbf{b}_\parallel.$ Such a gauge is shown in (c) for different choices of parameters and for different bands.}
    \label{fig:Gauge fixing in the Haldane model}
\end{figure}

In cases where $k_\parallel$ does traverse both $\mathbf{K}_+$ and $\mathbf{K}_-,$ the gauge choices above cannot evade all the singularities simultaneously. Thus, we need to patch this specific case in a different way.  Both $\mathbf{K}_+ = (k_{\parallel,+},k_\perp^*)$ and $\mathbf{K}_-= (k_{\parallel,-},k_\perp^*)$ reside on some critical $k_\perp^*$ and run along $k_\parallel.$  For $k_\perp \neq k_\perp^*,$ either Gauge I or II will suffice. Exactly at $k_\perp^*,$ we employ a regularization to make sure that the wave function is differentiable along the $k_\parallel$ direction (and \textit{only} along the $k_\parallel$ direction) using the wave function in, for example, Gauge II. We divide the $k_\parallel$ interval into two disjoint sections $\mathcal{S}_1 = \left[0, k_\parallel^*\right)$ and $\mathcal{S}_2 = \left[ k_\parallel^*, \frac{2\pi}{a_\parallel}\right),$ where $k_\parallel^*$ (which is either $k_{\parallel,+}$ or $k_{\parallel,-}$) is where the singularity resides in Gauge II. Then, we define
\begin{equation}
\begin{split}
    v_\pm^\text{(III)}(k_\parallel, k_\perp^*)&= \begin{cases}
    + v_\pm^\text{(II)}(k_\parallel, k_\perp^*) e^{ik_\parallel a_\parallel/2}, & \text{for } k_\parallel \in \mathcal{S}_1 \\
     - v_\pm^\text{(II)}(k_\parallel, k_\perp^*) e^{ik_\parallel a_\parallel/2}, & \text{for } k_\parallel \in \mathcal{S}_2
  \end{cases},\\
  v_\pm^\text{(III)}(k_{\parallel,\pm}, k_\perp^*) &= \lim_{k_\parallel \rightarrow k_{\parallel, \pm}} v_\pm^\text{(III)}(k_\parallel, k_\perp^*).
\end{split}
\end{equation}
This choice is inspired by the parallel transport procedure of Bloch wave functions used to define hybrid Wannier functions \cite{Marzari1997Maximally,Marzari2012Maximally}. We have picked Gauge II to define Gauge III, but that choice is  arbitrary. We could have easily done the same using Gauge I or some other quasi-continuous gauge. The important point is the sign change going from $\mathcal{S}_1$ to $\mathcal{S}_2.$ The exponential $e^{ik_\parallel a_\parallel/2}$ has been inserted to ensure the correct boundary condition. This factor is smooth in $k_\parallel,$ and therefore cannot alter the regularity of the defined gauge.    We now show that Gauge III is indeed differentiable along $k_\parallel.$ It is enough to inspect the differentiability of Gauge III around $\mathbf{K}_\pm:$
\begin{equation}
\begin{split}
    \hat{h}_x(\mathbf{K}_\pm + \mathbf{k}) - i \hat{h}_y(\mathbf{K}_\pm + \mathbf{k})& = \frac{t_1a}{3|t_2|} \left( \pm k_x -i k_y\right) + \mathcal{O}(k^2),\\
    \hat{h}_z(\mathbf{K}_\pm + \mathbf{k}) &= \pm \text{sgn}(t_2)  \mp \text{sgn}(t_2)\frac{t_1^2a^2}{18 t_2^2} \left( k_x^2+k_y^2 \right) + \mathcal{O}(k^3), \\ 
    e^{-i(\mathbf{K}_\pm + \mathbf{k}) \cdot \boldsymbol{\tau}_1} &= 1-\frac{i k_y a}{\sqrt{3}} + \mathcal{O}(k^2).
\end{split}
\end{equation}
If $t_2 > 0$ ($t_1$ is always assumed positive without loss of generality), we have to smallest order in $\mathbf{k}$
\begin{equation}
    \begin{split}
        v_+^\text{(II)}(\mathbf{K}_+ + \mathbf{k}) &= \begin{pmatrix}
            \frac{k_x-ik_y}{|\mathbf{k}|} \\
            \frac{|t_1|a|\mathbf{k}|}{6|t_2|}
        \end{pmatrix}, \\
        v_+^\text{(II)}(\mathbf{K}_- + \mathbf{k}) &= \begin{pmatrix}
            -\frac{a t_1 (k_x+i k_y)}{|t_2| \sqrt{36-\frac{a^2 t_1^2 }{t_2^2}|\mathbf{k}|^2}} \\
            \frac{1}{6} \sqrt{36-\frac{a^2 t_1^2 }{t_2^2}|\mathbf{k}|^2}
        \end{pmatrix},\\
        v_-^\text{(II)}(\mathbf{K}_+ + \mathbf{k}) &= \begin{pmatrix}
            \frac{a t_1 (k_x-i k_y)}{|t_2| \sqrt{36-\frac{a^2 t_1^2 }{t_2^2}|\mathbf{k}|^2}} \\
            -\frac{1}{6} \sqrt{36-\frac{a^2 t_1^2 }{t_2^2}|\mathbf{k}|^2}
        \end{pmatrix}, \\    v_-^\text{(II)}(\mathbf{K}_- + \mathbf{k}) &= \begin{pmatrix}
            \frac{-k_x-ik_y}{|\mathbf{k}|} \\
            -\frac{|t_1|a|\mathbf{k}|}{6|t_2|}
        \end{pmatrix}.\\
    \end{split}
\end{equation}
These eigenvectors are \textit{not} properly normalized because the Taylor expansion does not preserve the norm. This is unimportant at the moment since we only care about differentiability here. Clearly, $v_+^\text{(II)}(\mathbf{K}_- + \mathbf{k}) $ and $v_-^\text{(II)}(\mathbf{K}_+ + \mathbf{k}) $ are differentiable at $\mathbf{k} = \mathbf{0}.$ However, $v_+^\text{(II)}(\mathbf{K}_+ + \mathbf{k}) $ and $v_-^\text{(II)}(\mathbf{K}_- + \mathbf{k}) $ are \textit{not} differentiable: the upper, $A$ component presents a $2\pi$ phase singularity around $\mathbf{k} = \mathbf{0}$ while the lower, $B$ component contains a cusp singularity due to the absolute value. Both of these singularities are corrected along the $k_\parallel$ direction by alternating the sign going from ``left" to ``right" of the singularity. The analysis for $t_2 <0$ is almost exactly the same; so we shall skip it. As an example, we consider $\boldsymbol{\mathcal{E}} = \mathcal{E}\hat{x}.$ Then, we have $\mathbf{b}_\parallel = 2 \mathbf{b}_1 - \mathbf{b}_2 = \left(4\pi/a,0 \right),$ $\mathbf{b}_\perp = -\frac{1}{2}\mathbf{b}_2 = -\frac{2\pi}{\sqrt{3}a}\left(0,1\right),$ $a_\parallel = a/2,$ and $a_\perp = \sqrt{3}a.$ We choose the Brillouin zone to be $k_\parallel \times k_\perp \in  \left[0, \frac{4\pi}{a} \right) \times \left[0, - \frac{2\pi}{\sqrt{3}a} \right). $ The Dirac points are located at $(k_{\parallel,+}^*,k_\perp^*) = \left(\frac{4\pi}{3a},0 \right)$ and  $(k_{\parallel,-}^*,k_\perp^*) = \left(\frac{8\pi}{3a},0 \right).$ A differentiable gauge chosen along the critical line $k_\perp^*=0$ containing both Dirac cones is shown in Fig. \ref{fig:Gauge fixing in the Haldane model}. There, we show the real and imaginary parts of both the lower and upper components of the energy eigenvectors for positive and negative $t_2.$ As one can visually inspect, all of these components are without any discontinuity or kink and hence are differentiable. Also, they are periodic in $k_\parallel,$ i.e. the values at $k_\parallel = 0$ match those at $k_\parallel = 4\pi/a.$ Thus, we can define a Berry connection along $k_\parallel$ using such a gauge. The above considerations demonstrate that one can always choose a differentiable gauge along one direction, which, in our problem, is set by the electric field.

\begin{figure}
    \centering
    \includegraphics[width=6.5in]{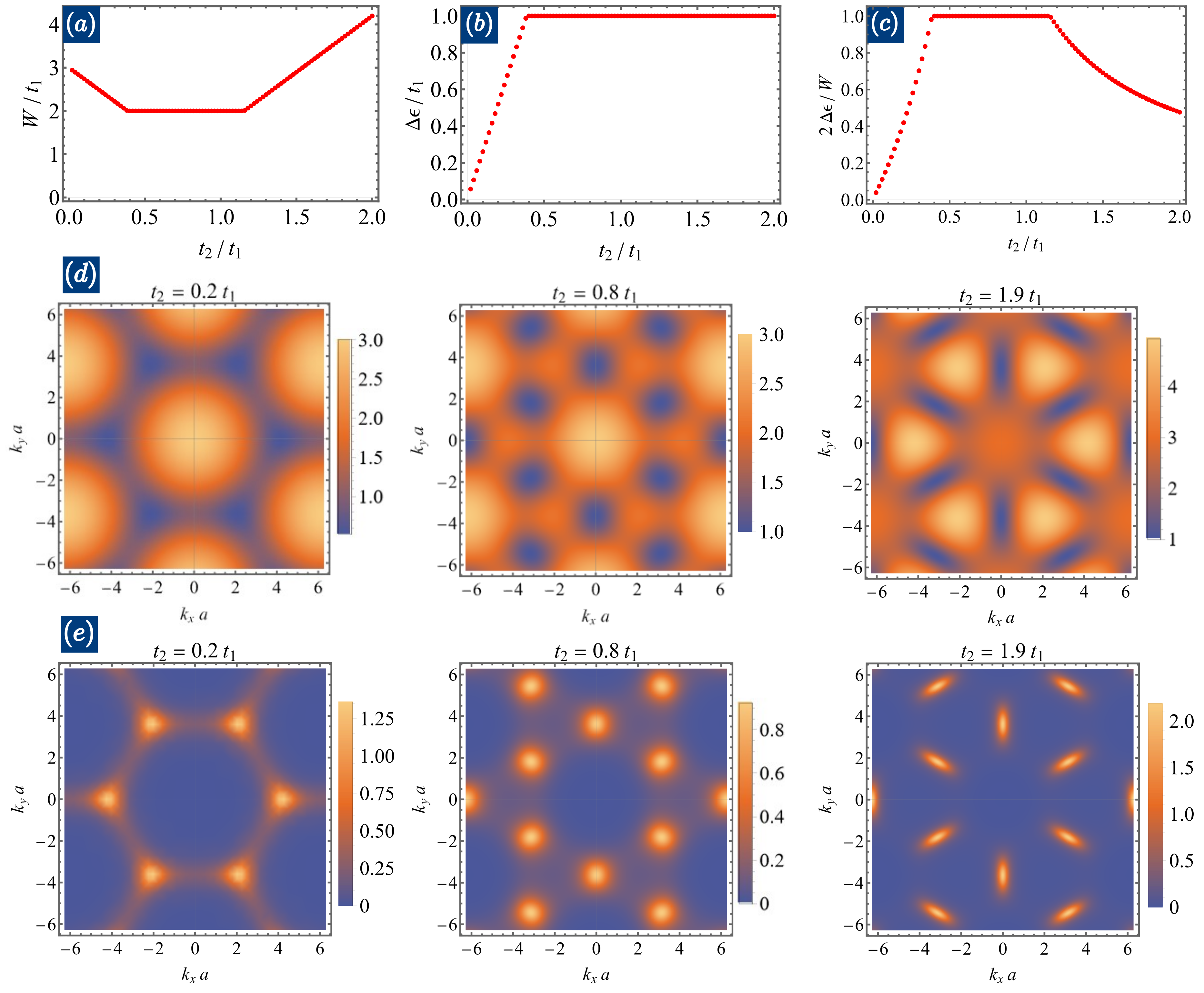}
    \caption{\textbf{Energetics and Berry curvature of the Haldane model.} Bandwidth (a), band gap (b), and gap-to-bandwidth ratio (c) of the Haldane model as a function of $t_2.$ Energy (d) and Berry curvature (e) of the valence band for three different sets of parameters.   }
    \label{fig:haldane2}
\end{figure}

Next, we consider the energetics of the Haldane model as a function of $t_2/t_1.$ The numerically-calculated bandwidth, band gap, and band-gap-to-bandwidth ratio are plotted in Fig. \eqref{fig:haldane2}(a,c). Here, since both the conduction and valence bands have the same bandwidth, we only measure the bandwidth of one. For small $t_2,$ the bandwidth decreases from $3t_1$ to about $2t_1.$ It then plateaus for intermediate values of $t_2/t_1.$ After that, the bandwidth increases with increasing $t_2/t_1.$ The band gap is zero at $t_2 = 0,$ increases linearly with $t_2,$ and plateaus around $\Delta \varepsilon/t_1=1$ upon reaching some critical $t_2.$ As a result, there is an interval in $t_2$ in which the band-gap-to-bandwidth ratio is optimal. The evolution of the $k$-space energy of the valence band as a function of $t_2$ is shown in Fig. \ref{fig:haldane2}(d). As $t_2$ increases, the band maxima shift from $\Gamma$ to $K_\pm.$ The evolution of the Berry curvature distribution is also shown in Fig. \ref{fig:haldane2}(e), where we observe the migration of the Berry curvature peaks from the zone corners $K_\pm$ at small $t_2$ to the $M$ points at large $t_2.$ Heuristically, the Berry curvature is concentrated near band minima.

\section{Assumed Occupation Function}
\label{sec: Assumed Occupation Function}

To calculate optical absorption, we only need matrix elements where $m \neq m'.$ In this case, shifting both $n$ and $n'$ by the same amount does not change the matrix elements and the difference of the associated energies $E_{n',m'}(k_\perp)-E_{n,m}(k_\perp) = E_{n'+\ell,m'}(k_\perp)-E_{n+\ell,m}(k_\perp)$. Also, we shall choose the occupation function in such a way that it respects translational symmetry along the direction of the electric field. We assume that the electron density is uniform on the unit cell scale along $\mathbf{b}_\parallel$ even in the presence of an electric field because the system is connected to metallic leads that can deposit and withdraw electrons on the chain \cite{Tsu1975Hopping,dohler1975new,Shon1996Hopping, rott1997hopping,Wacker1998Quantum, rott1999self, wacker2002semiconductor}. This choice of electron density is also sensible from a symmetry point-of-view. Even though the Hamiltonian does not manifestly respect translational symmetry due to our choice of gauge, a constant time-independent electric field should look the same at every point in space. This translational symmetry is manifest in the temporal gauge where $\boldsymbol{\mathcal{A}}(t) = \boldsymbol{\mathcal{E}} t.$ Furthermore, although probably not fully consistent, we continue to approximate the occupation function by Fermi-Dirac statistics supplemented by a spatially-dependent chemical potential: $\mu_{n,m}(k_\perp) = \mu + e \mathcal{E}\bar{\mathbf{r}}_{\parallel,n,m}(k_\perp),$ where $\bar{\mathbf{r}}_{\parallel,n,m}(k_\perp) = \bra{n,m,k_\perp} \hat{\mathbf{r}}_\parallel \ket{n,m,k_\perp}$ is the average position of the wave function. So the occupation at each site is 
\begin{equation}
    f\left[E_{n,m}(k_\perp)\right] = \frac{1}{1+e^{\beta\left[E_{n,m}(k_\perp)-\mu - e \mathcal{E}\bar{\mathbf{r}}_{\parallel,n,m}(k_\perp)\right] }},
\end{equation}
where $\beta$ is inverse temperature. The chemical potential $\mu$ is fixed by the  number density.   It is clear that this choice of chemical potential respects $f\left[E_{n,m}(k_\perp)\right] = f\left[E_{n+\ell,m}(k_\perp)\right]$ for all $\ell$ because $E_{n,m}(k_\perp)- e \mathcal{E}\bar{\mathbf{r}}_{\parallel,n,m}(k_\perp)=E_{n+\ell,m}(k_\perp)- e \mathcal{E}\bar{\mathbf{r}}_{\parallel,n+\ell,m}(k_\perp).$ The use of the Fermi-Dirac equilibrium occupation function does \textit{not} imply that we are dealing with an equilibrium system. In fact, the situation at hand is far from equilibrium in the limit of large electric field. With these assumptions, we can write the \textit{net} optical absorption as 
\begin{equation}
\label{eq: absorption coefficient 2 SM}
    \alpha (\omega)= \frac{\alpha_0\hbar \omega}{e^2N_\perp  }   \sum_{k_\perp,n',m\neq m'}  \abs{\bra{n',m',k_\perp} \hat{d} \ket{0,m,k_\perp}}^2 \delta \left( E_{n',m'}(k_\perp)-E_{0,m}(k_\perp)-\hbar \omega \right) \lbrace f\left[E_{0,m}(k_\perp) \right] -f\left[E_{n',m'}(k_\perp) \right] \rbrace,
\end{equation}
where $\alpha_0 = \pi e^2/n_r c \epsilon_0a_\parallel a_\perp  \hbar$ is an overall constant. Here, state $(0,m)$ is occupied and state $(n',m')$ is unocuppied.

\section{Optical Absorption}
\label{sec: Optical Absorption}

We consider shining monochromatic laser light at normal incidence to a superlattice material. The Hamiltonian is now modified to include a time-dependent term
\begin{equation}
    \hat{\mathcal{H}}_\text{rad}(t) = e \boldsymbol{\mathcal{E}}_\text{rad}(t) \cdot \hat{\mathbf{r}},
\end{equation}
where $\boldsymbol{\mathcal{E}}_\text{rad}(t)$ is the radiation field, which, unlike the in-plane static field, we assume to be weak. We choose the propagation in the $z$-direction and the superlattice material is placed on the $x$-$y$ plane. Therefore, we write the radiation field as 
\begin{equation}
    \boldsymbol{\mathcal{E}}_\text{rad}(t) = \mathfrak{Re} \left[\boldsymbol{\mathcal{E}}_\text{rad} e^{ik_zz - i \omega t}  \right],
\end{equation}
where $\boldsymbol{\mathcal{E}}_\text{rad} = \mathcal{E}_{\text{rad},\parallel}\hat{b}_\parallel + \mathcal{E}_{\text{rad},\perp} \hat{b}_\perp$ is a \textit{complex} two-dimensional vector that encodes the amplitude and polarization of the radiation field. By fixing the overall phase to be zero, we choose the parallel component to be purely real $\mathcal{E}_{\text{rad},\parallel}$ and write the perpendicular component explicitly with a phase $\phi_\text{rad}$ as $\mathcal{E}_{\text{rad},\perp}e^{i\phi_\text{rad}}.$ Then, we have 
\begin{equation}
\begin{split}
        \boldsymbol{\mathcal{E}}_\text{rad} &= \mathcal{E}_{\text{rad}}\left(n_{\text{rad},\parallel},n_{\text{rad},\perp}e^{i\phi_\text{rad}} \right) = \mathcal{E}_\text{rad} \mathbf{n}_\text{rad}, \\
        \mathcal{E}_{\text{rad}} &= \sqrt{\mathcal{E}_{\text{rad},\parallel}^2+\mathcal{E}_{\text{rad},\perp}^2}, \\
        n_{\text{rad},i} &= \frac{\mathcal{E}_{\text{rad},i}}{\mathcal{E}_{\text{rad}}}, \quad \mathbf{n}_\text{rad} \cdot \mathbf{n}_\text{rad}^* = 1.
\end{split}
\end{equation}
The Hamiltonian is then 
\begin{equation}
    \hat{\mathcal{H}}_\text{rad}(t) = \frac{e\mathcal{E}_\text{rad}}{2} \left( n_{\text{rad},\parallel} \hat{\mathbf{r}}_\parallel + n_{\text{rad},\perp} e^{i\phi_\text{rad}} \hat{\mathbf{r}}_\perp \right)e^{-i\omega t} + \frac{e\mathcal{E}_\text{rad}}{2} \left( n_{\text{rad},\parallel} \hat{\mathbf{r}}_\parallel + n_{\text{rad},\perp} e^{-i\phi_\text{rad}} \hat{\mathbf{r}}_\perp \right)e^{+i\omega t}.
\end{equation}
To simplify, we define a \textit{complex} dipole operator $\hat{d} = e \left(n_{\text{rad},\parallel} \hat{\mathbf{r}}_\parallel + n_{\text{rad},\perp} e^{i\phi_\text{rad}} \hat{\mathbf{r}}_\perp\right)$ to write
\begin{equation}
    \hat{\mathcal{H}}_\text{rad}(t) = \frac{\mathcal{E}_\text{rad}}{2} \hat{d}e^{-i\omega t} + \frac{\mathcal{E}_\text{rad}}{2} \hat{d}^\dagger e^{+i\omega t}.
\end{equation}
For linearly-polarized light, $\phi_\text{rad} = 0.$ For circularly-polarized light, $\phi_\text{rad} = \pm \pi/2$ and $n_{\text{rad},i} = \frac{1}{\sqrt{2}}.$ In general, we have elliptically-polarized light. Because $\mathcal{E}_\text{rad}$ is considered small (relative to some characteristic energy scale of the problem), we can apply time-dependent perturbation theory to study transitions between (quasi)stationary states. To lowest order, the transition rate from state $a$ to state $b$ with $E_b> E_a$ is given by Fermi's golden rule
\begin{equation}
    R_{a\rightarrow b} = \frac{2\pi}{\hbar} \frac{\mathcal{E}_\text{rad}^2}{4} \abs{\bra{b} \hat{d} \ket{a}}^2 \delta \left( E_b-E_a-\hbar \omega \right) f(E_a) \left[ 1-f(E_b)\right],
\end{equation}
where $f(E)$ is the occupation of a state at energy $E.$ The reverse process occurs with rate 
\begin{equation}
    R_{b\rightarrow a} = \frac{2\pi}{\hbar} \frac{\mathcal{E}_\text{rad}^2}{4} \abs{\bra{a} \hat{d}^\dagger \ket{b}}^2 \delta \left( E_a-E_b+\hbar \omega \right) f(E_b) \left[ 1-f(E_a)\right].
\end{equation}
Therefore, the \textit{net absorption rate per unit volume} is defined as $R_{a\rightarrow b} - R_{b\rightarrow a}$ \cite{chuang2012physics}
\begin{equation}
    R = \frac{1}{V} \sum_{a}\sum_{b}\frac{2\pi}{\hbar} \frac{\mathcal{E}_\text{rad}^2}{4} \abs{\bra{b} \hat{d} \ket{a}}^2 \delta \left( E_b-E_a-\hbar \omega \right)\left[ f(E_a) -f(E_b) \right],
\end{equation}
where $V$ is the volume of the absorptive material. This factor has been inserted to make $R$ intensive since the sum grows with volume (it may look like there are two sums that grow together like $V^2$, but this is not true since the occupation functions and matrix elements together limit the domain of the sums). The $\delta$ function enforces $E_b-E_a = \hbar \omega,$ which, for positive $\omega>0$, requires the transition to be from lower energy to higher energy. $R$ has units of inverse time (frequency) per unit volume (which is area for two dimensions and length for one dimension).  The absorption coefficient $\alpha$ is defined as the ratio of absorbed photons to incident photons
\begin{equation}
    \alpha(\omega) = \frac{\text{number of absorbed photons per unit time per unit volume}}{\text{number of incident photons per unit time per unit area}}.
\end{equation}
The number of incident photons can be found using the time-averaged Poynting vector. This gives the intensity, which is power per unit area:
\begin{equation}
    \mathcal{I} = \frac{n_rc \epsilon_0 \mathcal{E}_\text{rad}^2}{2} ,
\end{equation}
where $n_r$ is the refractive index of the medium (the dependence of which on $\omega$ is assumed weak, as usual, and henceforth neglected), $\epsilon_0$ is the vacuum permittivity, and $c$ is the speed of light. The number of photons incident per unit time per unit area is therefore $\mathcal{I} /\hbar \omega.$  Consequently, the absorption coefficient is given by 
\begin{equation}
\label{eq: absorption coefficient}
    \alpha (\omega)= \frac{\pi}{n_r c \epsilon_0 }  \left[ \frac{\omega}{V} \sum_{a}\sum_{b}  \abs{\bra{b} \hat{d} \ket{a}}^2 \delta \left( E_b-E_a-\hbar \omega \right)\left[ f(E_a) -f(E_b) \right] \right].
\end{equation}
The absorption coefficient has dimension $\left[ \text{length} \right]^{2-D},$ where $D$ is the dimension of the absorptive material. We comment in passing that the combination of fundamental constants $\pi/n_r c \epsilon_0$ defines the units of Fig. 4 of the main text. When $D = 3,$ the absorption coefficient has units of inverse length. When $D = 2,$ the absorption coefficient is dimensionless. Curiously, when $D =1,$ the absorption coefficient has dimensions of length. As expected, $\alpha$ is independent of the field amplitude.  In numerical calculation, we broaden the peaks by replacing the $\delta$ function with its Lorentzian approximation
\begin{equation}
    \delta \left( E_b-E_a-\hbar \omega \right) \rightarrow \frac{1}{\pi} \frac{\eta}{ \eta^2 + \left( E_b-E_a-\hbar \omega \right)^2},
\end{equation}
where $\eta$ is a broadening factor. This broadening accounts for some scattering that is inevitably present in an experiment.

\section{Evaluation of Various Position Expectation Values}
\label{sec: Evaluation of Various Position Expectation Values}

To compute the optical absorption, we need various position expectation values. Recalling that $\Psi(\mathbf{k}) = \mathcal{L}^\dagger(\mathbf{k}) \Phi(\mathbf{k})$ and $\Phi(\mathbf{k}) = \mathcal{V}(\mathbf{k})^\dagger \Omega(\mathbf{k}),$ we can write $\Psi_{n,m}(k_\parallel,k_\perp) =  \sum_{n',m'} c_{n',m'}^{(n,m)}(k_\perp)\psi_{n',m'}(k_\parallel,k_\perp).$ Thus, in the band basis, the eigenfunctions can be written explicitly as
\begin{equation}
    \Omega_{n,m}(k_\parallel,k_\perp) = \frac{1}{\sqrt{N_\parallel}} \mathcal{V}(k_\parallel, k_\perp) \exp \left(\frac{i}{e\mathcal{E}} \int_0^{k_\parallel} \left[ \mathcal{D}(k_\parallel',k_\perp) - \bar{\mathcal{D}}(k_\perp)\right]dk_\parallel'\right)   \sum_{n',m'} c^{(n,m)}_{n',m'}(k_\perp)e^{-in'k_\parallel a_\parallel}  \mathbb{1}_{m'} .
\end{equation}
The associated states in the position basis are 
\begin{equation}
    \ket{n,m,k_\perp} = \frac{1}{\sqrt{N_\perp}} \sum_{\mathbf{r},\sigma} e^{ik_\perp \hat{b}_\perp \cdot \left( \mathbf{r}+ \boldsymbol{\tau}_\sigma \right)} \left[ \frac{1}{\sqrt{N_\parallel}} \sum_{k_\parallel} \left[ \Omega_{n,m}(k_\parallel,k_\perp) \right]_\sigma e^{i k_\parallel \hat{b}_\parallel \cdot \left(\mathbf{r}+ \boldsymbol{\tau}_\sigma \right)}  \right] \ket{\mathbf{r},\sigma},
\end{equation} 
where $\ket{\mathbf{r},\sigma} = \hat{c}^\dagger_{\mathbf{r},\sigma} \ket{0}.$ The parallel position expectation value is 
\begin{equation}
\label{eq: parallel position operator}
    \bra{n',m',k_\perp'} \hat{\mathbf{r}}_\parallel \ket{n,m,k_\perp} = \delta_{k_\perp',k_\perp} \left[ i \sum_{k_\parallel} \Omega_{n',m'}^\dagger(k_\parallel,k_\perp) \frac{\partial}{\partial k_\parallel}  \Omega_{n,m}(k_\parallel,k_\perp) \right].
\end{equation}
We observe that because $\Omega_{n+\ell,m}(k_\parallel,k_\perp) = e^{-i \ell k_\parallel a_\parallel}\Omega_{n,m}(k_\parallel,k_\perp),$ we have following relations
\begin{equation}
    \bra{n'+\ell', m',k_\perp} \hat{\mathbf{r}}_\parallel \ket{n+\ell,m,k_\perp} = i \sum_{k_\parallel} e^{i(\ell'-\ell) k_\parallel a_\parallel} \Omega_{n',m'}^\dagger(k_\parallel,k_\perp)  \frac{\partial}{\partial k_\parallel}  \Omega_{n,m}(k_\parallel,k_\perp) + \ell a_\parallel \delta_{n,n'}\delta_{m,m'}\delta_{\ell,\ell'}.
\end{equation}
By using the band basis, Eq. \eqref{eq: parallel position operator} has a nice analytic form of a Berry connection along the $k_\parallel$ direction. However, there is an alternative method to compute the parallel position expectation value that may be more computationally convenient. In this approach, we write the parallel position operator directly in the ladder basis
\begin{equation}
\label{eq: parallel position operator 2}
    \begin{split}
        \mathbf{r}_\parallel &= \exp \left[ -\frac{i}{e\mathcal{E}} \int_0^{k_\parallel} \mathcal{D}(k'_\parallel, k_\perp) dk_\parallel' \right] \left( \mathcal{A}_\parallel (k_\parallel, k_\perp) + \mathbb{1} i \frac{\partial}{\partial k_\parallel}\right) \exp \left[ +\frac{i}{e\mathcal{E}} \int_0^{k_\parallel} \mathcal{D}(k'_\parallel, k_\perp) dk_\parallel' \right] \\
        &= - \frac{1}{e\mathcal{E}} \mathcal{D}(k_\parallel, k_\perp) +  \mathcal{A}_\text{L}^\text{full} (k_\parallel, k_\perp) + \mathbb{1} i \frac{\partial}{\partial k_\parallel},
    \end{split}
\end{equation}
where $\mathcal{A}_\text{L}^\text{full}$ is the full non-Abelian Berry connection matrix in the ladder representation (which differs from $\mathcal{A}_\text{L}$ since this only contains the off-diagonal elements). Using Eq. \eqref{eq: parallel position operator 2}, the expectation values are computed simply as inner products with the parallel position operator matrix. Though both are formally equivalent, the advantage of using Eq. \eqref{eq: parallel position operator 2} over Eq. \eqref{eq: parallel position operator} lies in the fact that the former does not require the evaluation of a numerical derivative of $\Omega_{n,m}(k_\parallel,k_\perp)$ that may necessitate a fine mesh in order to capture possible fluctuations of the wave function in $k_\parallel.$

We will also need the perpendicular position expectation value. This calculation requires much more care since the energy eigenstates are plane waves in that direction, so their position expectation values might not be well-defined. We restrict our attention to calculating only the expectation values between \textit{non-degenerate} states. In general, we cannot use $i \sum_{k_\parallel} \Omega_{n',m'}^\dagger(k_\parallel,k_\perp) \frac{\partial}{\partial k_\perp}  \Omega_{n,m}(k_\parallel,k_\perp)$ because the wave functions have not necessarily been chosen to be continuous in that direction. More fundamentally, if the bands are topologically obstructed, a gauge cannot be chosen that is smooth both in $k_\parallel$ and $k_\perp.$ To circumvent this problem, we employ a familiar technique to calculate this position expectation value by replacing $\hat{\mathbf{r}}_\perp$ with the commutation relation $\left[ \hat{\mathbf{r}}_\perp, \hat{\mathcal{H}}\right]$ in \cite{vanderbilt2018berry}
\begin{equation}
    \bra{n',m',k_\perp'} \hat{\mathbf{r}}_\perp \ket{n,m,k_\perp} = \frac{\bra{n',m',k_\perp'} \left[\hat{\mathbf{r}}_\perp, \hat{\mathcal{H}}\right] \ket{n,m,k_\perp} }{E_{n,m}(k_\perp)-E_{n',m'}(k_\perp)}.
\end{equation}
The commutator is commonly evaluated  as a derivative of the Hamiltonian in an appropriate representation. In our case, $\left[ \hat{\mathbf{r}}_\perp, \hat{\mathcal{H}} \right] = \left[ \hat{\mathbf{r}}_\perp, \hat{\mathcal{H}}_0 \right],$ where $\hat{\mathcal{H}}_0$ is the Hamiltonian without the electric field, because the part containing the electric field is proportional to $\hat{\mathbf{r}}_\parallel$ that must commute with $\hat{\mathbf{r}}_\perp.$ In the orbital basis in momentum space, we have    $\left[\hat{\mathbf{r}}_\perp ,\hat{\mathcal{H}}_0 \right] = i \partial_{ k_\perp} \mathcal{H}_0(k_\parallel,k_\perp).$ Consequently, we have
\begin{equation}
\label{eq: perpendicular position operator}
   \bra{n',m',k_\perp}\hat{\mathbf{r}}_\perp\ket{n,m,k_\perp} = \frac{i }{E_{n,m}(k_\perp)-E_{n',m'}(k_\perp)}\sum_{k_\parallel} \Omega_{n',m'}^\dagger(k_\parallel, k_\perp)  \frac{\partial \mathcal{H}_{0}(k_\parallel, k_\perp)}{\partial k_\perp}  \Omega_{n,m}(k_\parallel, k_\perp).
\end{equation}
Again, we have the following relation
\begin{equation}
\begin{split}
    \bra{n'+\ell',m',k_\perp}\hat{\mathbf{r}}_\perp\ket{n+\ell,m,k_\perp} &= \frac{i }{E_{n,m}(k_\perp)-E_{n',m'}(k_\perp)+(\ell-\ell')e\mathcal{E}a_\parallel} \times \\
    &\times \sum_{k_\parallel} e^{i(\ell'-\ell)k_\parallel a_\parallel}\Omega_{n',m'}^\dagger(k_\parallel, k_\perp)  \frac{\partial \mathcal{H}_{0}(k_\parallel, k_\perp)}{\partial k_\perp}  \Omega_{n,m}(k_\parallel, k_\perp).
\end{split}
\end{equation}
We note that Eq. \eqref{eq: perpendicular position operator} would also work for $\hat{\mathbf{r}}_\parallel$ by replacing $\partial_{k_\perp}$ with $\partial_{k_\parallel},$ which can serve as a consistency check on these formulas. However, Eq. \eqref{eq: parallel position operator} is more general since it also works for computing the average parallel position of a single state.

\twocolumngrid

\bibliography{references.bib}

\end{document}